\documentclass[11pt]{article}
\usepackage[a4paper,margin=25mm,headheight=14pt]{geometry}
\usepackage[T1]{fontenc}
\usepackage[utf8]{inputenc}
\usepackage{lmodern}
\usepackage{amsmath,amssymb,mathtools,bm,slashed,mathrsfs}
\usepackage{longtable,array,tabularx}
\usepackage{tikz}
\usetikzlibrary{arrows.meta, positioning, calc}
\usepackage[numbers,sort&compress]{natbib}
\usepackage{graphicx,xcolor,microtype,enumitem,fancyhdr}
\usepackage[unicode,colorlinks=true,linkcolor=blue!45!black,citecolor=blue!45!black,urlcolor=blue!50!black]{hyperref}
\hypersetup{pdftitle={Survey of Mother Theories, I: Bosonic Fuzzy Sphere CFTs},pdfauthor={Mykola Dedushenko}}
\newcommand{\doi}[1]{\href{https://doi.org/#1}{doi:\nolinkurl{#1}}}
\allowdisplaybreaks[2]
\numberwithin{equation}{section}
\newcommand{\dd}{\mathrm d}
\newcommand{\ii}{\mathrm i}
\newcommand{\R}{\mathbb R}
\newcommand{\Z}{\mathbb Z}

\newcommand{\cL}{\mathcal L}
\newcommand{\cH}{\mathcal H}
\newcommand{\Tr}{\operatorname{Tr}}
\newcommand{\tr}{\operatorname{tr}}
\newcommand{\LLL}{\mathrm{LLL}}
\newcommand{\HLL}{\mathrm{HLL}}
\newcommand{\ep}{\mathrm{ep}}
\newcommand{\cK}{\mathcal{K}}
\title{\bfseries Survey of Mother Theories, I:\\[4pt] Bosonic Fuzzy Sphere CFTs}
\author{Mykola Dedushenko\\[4pt]
\small Center for Mathematics and Interdisciplinary Sciences, Fudan University\\
\small Shanghai, 200433, China\\
\small Shanghai Institute for Mathematics and Interdisciplinary Sciences (SIMIS)\\
\small Shanghai, 200433, China}
\date{October 2026}
\begin{document}
\maketitle
\begin{abstract}
We study non-gauge mother theories for the known fuzzy sphere Hamiltonians realizing bosonic 3D CFTs. In each construction, neutral mediators couple to charged fields in a magnetic background on $S^2\times \R$. Eliminating the mediators and higher Landau levels reproduces the Hamiltonian on the lowest Landau level, while eliminating the charged degrees of freedom gives the neutral effective theory. We explain the details for the following 3D examples: Ising and its deformations, easy-plane models, free scalars, $O(n)$ and cubic models, three-state Potts, as well as $SO(5)$, $O(4)$, and $Sp(n)$ models, whose EFT is a nonlinear sigma model with level-one WZ term. We find the $z=2$ criticality in the easy-plane case (whose Lorentz symmetry is obstructed by a certain Berry connection term in the effective action), and relativistic CFT in all other cases. Scalar mediators, also called Hubbard-Stratonovich fields, suffice for all these examples. Some further speculations are collected in the Outlook.
\end{abstract}
\newpage
\tableofcontents

\section{Introduction}\label{sec:introduction}

The fuzzy sphere construction replaces spatial functions on $S^2$ by a finite orbital algebra while preserving rotations~\cite{Madore1992,Haldane1983,GrosseKlimcikPresnajder1996}. It is by itself a quantum system (quantization of $S^2$) that can be obtained in a variety of ways (since quantization methods are abundant), but a particularly useful realization for us is as the lowest Landau level (LLL) of a charged $(2+1)D$ particle on $S^2\times \R$ in the magnetic field of uniform flux $Q$ through $S^2$. In the recent realizations of $(2+1)D$ conformal field theories (CFTs) as quantum critical points of many-body systems on the fuzzy sphere \cite{ZhuIsing}, charged particles occupy the LLL, and their neutral collective modes supply the continuum CFT fields. The paradigmatic example of Ising model construction in \cite{ZhuIsing} established that this regulator can resolve the operator spectrum of an interacting three-dimensional conformal field theory. Subsequent constructions reach vector and rotor models, free scalars, discrete symmetry transitions, topological order parameter theories, fermionic critical points, and several gauge-theoretic examples. Each existing construction first states what particles to fill the sphere with, which is often chosen based on the expected symmetries and field content of the target CFT, provides a Hamiltonian with local interactions on $S^2$ (subsequently projected to the LLL) respecting those symmetries, and tunes it to the desired CFT, which in most cases means tuning to the quantum critical point. An exception to the latter is the realization of free scalars as pseudo-Goldstone modes \cite{HeFreeScalar,GuoON}, for which, instead of tuning to criticality, one stays inside the spontaneous symmetry breaking (SSB) phase, realizing the free scalar as a Goldstone particle (deformed by the conformal mass on curved $S^2\times \R$ background, which is the reason for ``pseudo'').

The companion Letter \cite{MotherPRL} outlines a new perspective on this approach, putting it within the toolbox of Quantum Field Theory (QFT) and revealing its hidden effective field theory (EFT) origins. In particular, rather than viewing the emergent CFT as a miracle of universality, it describes these Hamiltonians as descendants of local \emph{mother theories} on $S^2\times\R$, i.e. as low energy truncations of the full QFT. A mother theory contains gapped charged fields $\Psi$, coupled to the external monopole connection of flux $Q$, and gapless neutral mediators $\Phi$. Integrating out the gapped charged fields gives a conventional neutral EFT tunable to the desired CFT point. Integrating out the mediators instead gives interactions among the charged particles; their LLL projection gives the fuzzy Hamiltonian:
\begin{center}
\begin{tikzpicture}[
	basebox/.style={
		draw,
		rounded corners,
		minimum height=0.8cm,
		inner ysep=1.5pt,
		align=center
	},
	motherbox/.style={
		basebox,
		minimum width=0pt,
		inner xsep=5pt
	},
	lowerbox/.style={
		basebox,
		minimum width=3.0cm,
		inner xsep=6pt
	},
	arrow/.style={
		-{Latex[length=2mm]},
		thick,
		rounded corners=1.5pt
	},
	arrowlabel/.style={
		font=\scriptsize,
		inner sep=1pt,
		fill=white
	}
	]
	\node[motherbox] (mother) at (0.4,1.55)
	{$\cL_{\rm mother}[\Psi,\Phi]$};
	\node[lowerbox] (fuzzy) at (-2.35,0.3)
	{$\cH_{\rm fuzzy}$};	
	\node[lowerbox] (eft) at (2.95,0.3)
	{$\cL_{\rm local\ EFT}
		\xrightarrow{\rm RG}
		{\rm CFT}$};
	\coordinate (lefttarget)
	at ($(fuzzy.north west)+(0.42cm,0)$);
	\coordinate (righttarget)
	at ($(eft.north east)+(-0.42cm,0)$);
	\coordinate (leftbend)
	at (lefttarget |- mother.south west);
	\coordinate (rightbend)
	at (righttarget |- mother.south east);
	\draw[arrow]
	(mother.south west)
	--
	node[pos=0.50, above=2pt, arrowlabel]
	{integrate out $\Phi$}
	node[pos=0.50, below=2pt, arrowlabel]
	{project to LLL}
	(leftbend)
	--
	(lefttarget);
	\draw[arrow]
	(mother.south east)
	--
	node[pos=0.50, above=2pt, arrowlabel]
	{integrate out $\Psi$}
	(rightbend)
	--
	(righttarget);
\end{tikzpicture}\end{center}
Then the same CFT emerges in the thermodynamic limit of $\cH_{\rm fuzzy}$, defined as $Q\to\infty$, $R\to\infty$, $Q/R^2={\rm fixed}$, where $R$ is the radius of $S^2$. In this treatment we also assume that the mass scale $m$ of $\Psi$ and the Landau level gap $\Delta_{\rm LL}$ are separated, $m\ll \Delta_{\rm LL}$, since otherwise the LLL projection would be intractable. A complete treatment of the model makes both integration routes explicit, thus making the interaction channels and critical tunings more transparent.

The Letter \cite{MotherPRL} explained the method using only two families of mother theories, Ising and easy-plane. In the current paper, the goal is to make progress with our program of identifying mother theories for all known fuzzy sphere Hamiltonians \cite{ZhuIsing,HanBilayer,topZhouSO52024,topChenPhase2024,topChen2024,topYang2026,topZhouHe2025,pc:YangPotts,FanLeeYang,ArguelloLeeYang,EliasLeeYang,HeFreeScalar,TaylorTricritical,GuoON,DeyOthree,DeyOtwo,StergiouCubic,fer:ZGH,fer:Voinea,fer:Tang,VoineaFanRegnaultPapic2025,ZhouWangHe2025,topHuffman2026}. Specifically, here we only apply the procedure to the models listed below, which are bosonic CFTs admitting the non-gauge mother theories. \emph{Non-gauge} means that the mother has no dynamical gauge fields. Some of the models, $SO(5)$, $O(4)$, and $Sp(n)$, are included through their non-gauge mothers (order parameter descriptions), even if proposed infrared dual gauge-theoretic descriptions exist. Cases requiring dynamical gauge fields in the mother will be treated separately. The Chern--Simons matter model~\cite{ZhouWangHe2025} and fractional-filling Ising constructions with a topological charge sector~\cite{VoineaFanRegnaultPapic2025} belong to that broader program; their additional sectors are not constructed here. For each family we specify the fields, mass or interaction matrices, filling, and mediator kernels. We compute the charged state response in the magnetic background and derive the interaction obtained in the opposite integration order. Where several published models have the same target CFT, we keep their different microscopic channels separate: a singlet--triplet rotor \cite{DeyOthree}, a Heisenberg bilayer \cite{HanBilayer}, and a real vector exchange model \cite{GuoON} are different mothers for the $O(3)$ criticality.

The magnetic background flux is essential to the argument. It explicitly breaks Lorentz invariance, but one might hope for its restoration in the IR since the background flux does not directly couple to the light neutral degrees of freedom. In the minimal easy-plane doublet mother considered in \cite{MotherPRL}, for example, the effective scalar potential is compatible with relativistic $O(2)$ Wilson--Fisher criticality. The occupied LLL, however, generates a term in the EFT with first order time derivative, i.e. the Berry term derived in \cite{MotherPRL}. Thus the actual saturation transition in that case had $z=2$ scaling; the Lorentz symmetry was \emph{not} restored. In every model we analyze, we must therefore check whether such a Berry term is present in order to know if the IR CFT can be relativistic. If the Berry term is present, it obstructs the restoration of Lorentz symmetry. In real vector and integer rotor mothers, a real occupied line will remove this Berry response. In quaternionic families the determinant line will be trivial, while a higher topological term remains. The analysis of Berry term therefore determines which neutral EFT is available before the precise infrared identification is made.

For the Ising, $O(n)$, and cubic families, the local mother theories connect the microscopic fuzzy sphere Hamiltonians \cite{ZhuIsing,HanBilayer,GuoON,DeyOthree,DeyOtwo,StergiouCubic} to EFTs with familiar renormalization group flows landing at the expected CFTs. For the three-dimensional Potts and deconfined models~\cite{pc:YangPotts,topZhouSO52024,topChenPhase2024,topChen2024,topYang2026,topZhouHe2025,topHuffman2026}, the existence or nature of a fixed point is itself part of the physical problem. There we derive the relevant scalar EFT and reproduce the published interaction channels. In the $Sp(n)$ families, relating an exact auxiliary field representation to a local order parameter theory also requires the charge and radial gaps to persist along the RG trajectory. The resulting EFT may describe a real fixed point, a multicritical point, or pseudocritical behavior; these possibilities are not settled by the matching done in this work. Similarly, a non-Gaussian mediator could reproduce a desired neutral sector while generating many-body interactions in the fuzzy Hamiltonian.

We mostly focus on constructing non-gauge mother theories of the known fuzzy sphere models for bosonic CFTs. However, several new constructions follow from the same calculations along the way. In the Outlook, we speculate about a Berry-balanced easy-plane mother, a quadratic Yukawa deformation generating a nonzero $\Z_3$ three-flip interaction, compact and projective mediators without elementary radial fields (which contaminate the pseudo-Goldstone spectrum in the free scalar example \cite{MotherPRL}), etc. The required kernels and limiting procedures are to be determined in each case.

Fermionic CFTs, which require including fermionic mediators (a natural generalization, compared to the case of bosonic mediators) will be studied in the upcoming work. Local fermionic CFTs are different from gauged Majorana and gauged super-Ising theories~\cite{fer:ZGH,fer:Voinea,fer:Tang}, which pose interesting challenges for this future work. We restrict attention to bulk models; defect theories, dynamical-gauge mothers, and further classification of fuzzy regulators lie outside this survey.

The catalog of theories considered here, along with their main features, is given below.\newpage
\begin{enumerate}
	\item \textbf{Ising}~\cite{ZhuIsing,MotherPRL}.	A Dirac doublet couples to a real Ising scalar and, optionally,	a density auxiliary. The resulting real scalar EFT has no Berry	term. Criticality is known numerically.
	
	\item \textbf{Tricritical Ising and Lee--Yang}%
	~\cite{TaylorTricritical,FanLeeYang,ArguelloLeeYang,EliasLeeYang}.
	Independent density and Ising kernels, with an imaginary Ising-odd source for Lee--Yang, give Gaussian tricritical and	imaginary cubic EFTs. The kernels reproduce the published interactions; infrared identifications rely on cited numerics.
	
	\item \textbf{Easy-plane family}~\cite{MotherPRL}.
	A charged doublet couples to a complex scalar. A nonzero Berry term gives $z=2$ saturation determined analytically. In the ordered phase, conformal pinning gives a free relativistic scalar.
	
	\item \textbf{He's free scalar}~\cite{HeFreeScalar}.
	Same as easy-plane model, plus an irrelevant deformation.
	
	\item \textbf{$O(n)$ models and integer rotors}%
	~\cite{GuoON,DeyOtwo,DeyOthree}.
	Dirac fermions --- a singlet and $n$ flavor components --- couple to $n$ order parameter scalars and density auxiliaries, giving a real $n$-component EFT. Filling and derivative interactions are specified for each Hamiltonian. No Berry response.
	
	\item \textbf{Heisenberg bilayer}~\cite{HanBilayer}.
	Four charged flavors with layer and spin indices couple to staggered and uniform spin mediators. The resulting $O(3)$ order parameter has no quadratic Berry response.
	
	\item \textbf{Cubic model}~\cite{StergiouCubic}.
	$O(3)$ rotor flavors couple to additional scalar auxiliaries for cubic-symmetric bilinears. Two independent quartic invariants give the cubic EFT; no Berry response. Criticality determined in the published work.
	
	\item \textbf{Free $O(n)$ scalars}~\cite{GuoON}.
	There are $n+1$ charged flavors, with density and scalar derivative auxiliaries. The EFT contains $n$ Goldstone coordinates with conformal pinning; massive conjugate modes generate the temporal kinetic term. No Berry term generated. Alternative EFT analysis presented.
	
	\item \textbf{Three-state Potts}~\cite{pc:YangPotts}.
	Three charged Dirac flavors couple to a complex clock mediator and a density auxiliary. The EFT has a $Z_3$ scalar potential with a cubic invariant and no quadratic Berry term. A real CFT is not established by the EFT calculation alone.
	
	\item \textbf{$SO(5)$ and $O(4)$}%
	~\cite{topZhouSO52024,topChenPhase2024,topChen2024,topYang2026}.
	Four Dirac flavors couple to five Clifford scalar mediators. A nonzero radial minimum permits an $S^4$ Wess--Zumino EFT; the $O(4)$ restriction gives theta-term $\theta=\pi$. Charge and radial gaps are required. No Berry term generated.
	
	\item \textbf{$Sp(n)$}~\cite{topZhouHe2025,topHuffman2026}.
	There are $2n$ charged Dirac flavors, with scalar mediators in specified matrix representations. The resulting EFT is a nonlinear sigma model of a quaternionic order parameter with a WZ term. No Berry term.
\end{enumerate}

\section{Mother theories in magnetic background: general strategy}\label{sec:framework}

\subsection{The two matching problems}
For each mother theory we ask three questions. What neutral EFT is obtained when the charged fields are eliminated in the specified magnetic background and filling on $S^2\times \R$? Can its couplings be tuned to the intended infrared theory? What interactions are obtained when the mediators are eliminated? The first requires a charged gap and a calculation of the response; the second requires the infrared dynamics of the resulting EFT. The third usually requires a simple Gaussian integration, the Hubbard--Stratonovich (HS) transformation, however, in some special situations the mediator (also called the HS field in \cite{MotherPRL}) may be necessarily non-Gaussian, which requires special treatment. Finally, in a slight variation of the procedure referred to as the ``reverse route'' in \cite{MotherPRL}, charged fields are assumed to be LLL-projected from the beginning, and the HS mediator used to decouple their interactions comes with a natural cutoff for its spherical harmonics as well. Then the third question is asked in the reversed order: what interaction channels in the mother theory action (that couple charged fields to the HS mediators) are required in order to reproduce the expected Hamiltonian on the LLL upon HS fields being integrated out.

Now we will provide more details. For each formula we have to be careful whether it is an exact identity after LLL projection, the leading terms in a local derivative expansion, or an infrared identification. For non-Gaussian (in particular, constrained) mediators, integration (that may be called the nonlinear HS transformation) involves connected correlators of all orders. Keeping only the two-body interactions often present in $\cH_{\rm fuzzy}$ then needs a separately motivated approximation. The latter comment matters especially for the topological and fermionic families.

\subsection{Flux, spin, and the cutoff}\label{sub:flux}
We use a background $U(1)$ connection $A$ of uniform flux $Q$,
\begin{equation}\label{eq:flux}
	Q=\frac{1}{2\pi}\int_{S_R^2}F\in\mathbb Z_{>0},\qquad \mathcal A=4\pi R^2.
\end{equation}
Depending on whether the charge $1$ field is scalar or spinor, set:
\begin{align}\label{eq:scalar-spinor}
	\text{Scalar:}\quad & N=Q+1,\cr
	\text{Spinor:}\quad & N=Q.
\end{align}
Then define:
\begin{equation}
	s=\frac{N-1}{2},\qquad \rho_0=\frac{N}{\mathcal A}.
\end{equation}
Ordinary spherical harmonics are normalized by $\int\dd A\,Y_{LM}^*Y_{L'M'}=\delta_{LL'}\delta_{MM'}$, and $Y_{00}=\mathcal A^{-1/2}$. We use the time Fourier convention $X(\tau)=\int(\dd\omega/2\pi)e^{-\ii\omega\tau}X(\omega)$. Scalar monopole sphere harmonics are similarly normalized, and the LLL eigenfunctions are denoted $u_m$.

The orbital number is $N$, like in the companion Letter, and stands for the dimension of LLL for a charge $+1$ field. The relative unit shift between scalars and spinors is due to the spin connection on $S^2$. Both for scalars and spinors, the LLL forms a spin-$s$ representation $V_s$ of the $SU(2)$ spatial rotations. The nonzero spatial Dirac eigenvalues and degeneracies are \cite{Dolan2003,JanteSchroers}
\begin{equation}\label{eq:diracLL}
	\lambda_n^2=\frac{n(Q+n)}{R^2},\qquad d_n=Q+2n,\qquad n\geq1,
\end{equation}
for each sign of $\lambda_n$. For scalars, the HLL covariant Laplacian eigenvalues and degeneracies are:
\begin{equation}
	\lambda_n^2 = \frac{j(j+1)-Q^2/4}{R^2}=\frac{n(Q+n+1)+Q/2}{R^2},\qquad j=\frac{Q}{2}+n,\qquad d_n=Q+2n+1.
\end{equation}
The one-unit shift between scalars and spinors matters for fermionic descendants and for precise finite size matching. Scalar monopole wavefunctions of flux $Q-1$ and Dirac zero modes at physical flux $Q$ have the same orbital representation after the spin connection is included; this is the dictionary used when comparing to spinless LLL Hamiltonians.

With a constant flavor mass eigenvalue $M$, the higher energies for spinors are $\pm\sqrt{M^2+\lambda_n^2}$, while for scalars they are $\sqrt{M^2 + \lambda_n^2 + \xi {\mathcal R}}$, where ${\mathcal R}$ is scalar curvature, and $\xi=0$ for minimal curvature coupling, while conformal coupling in 3D is $\xi=1/8$. For spinors the LLL is chiral (only one component of a spinor is non-zero), and we choose the chirality convention in which the LLL one-body Hamiltonian equals the flavor matrix $\mathsf M$; the opposite convention reverses its signs and occupied subspaces. In particular, for $\mathsf M>0$, the Dirac vacuum has empty LLL, while for $\mathsf M<0$, the LLL is filled. All filling statements below refer to this convention.

The orbital algebra and a general number-conserving bilinear are
\begin{equation}\label{eq:orbitalalgebra}
	\operatorname{End}(V_s)=\bigoplus_{L=0}^{N-1}V_L,\qquad n^T_{LM}=\sum_{mm',ab}c^\dagger_{ma}(\widehat Y_{LM})_{mm'}T_{ab}c_{m'b}.
\end{equation}
Our real-space densities use wavefunctions normalized in $\dd A$, so that $\sum_m|u_m(x)|^2=\rho_0$. We have to be mindful of these conventions when comparing with the existing literature that might use different normalizations.

The relevant EFT window is:
\begin{equation}\label{eq:hierarchy}
	E_{\rm neutral}\ll\Delta_c,\qquad E_{\rm retained},\ E_{\rm mix}\ll\Delta_{\rm LL},\qquad L\ll\sqrt Q,\qquad \ell_B=\frac{\sqrt2R}{\sqrt Q},
\end{equation}
where $\Delta_c\sim m$ is the charged excitation gap, $E_{\rm mix}$ measures matrix elements mixing Landau levels, and $\Delta_{\rm LL}$ is the appropriate excitation gap to those levels. For $|M|R\ll\sqrt Q$, $\Delta_{\rm LL}\sim\sqrt Q/R$. Interactions that mix Landau levels must be small compared to this gap. The algebraic maximum $L=N-1$ is much larger than the angular momenta described by a local derivative expansion, which requires $L\ll\sqrt Q$.

The thermodynamic limit used in the Introduction holds the magnetic field $B=Q/(2R^2)$ fixed. Thus $\rho_0=B/(2\pi)$ for spinor (or $\rho_0-\frac1{\mathcal A}=B/(2\pi)$ for scalar) and $\ell_B=B^{-1/2}$ remain fixed, while $R^{-1}\to0$. We choose the microscopic gaps and interaction strengths to satisfy \eqref{eq:hierarchy} before taking this limit. Increasing $Q$ at fixed $B$ does not itself increase the Landau gap; it extends the range between the finite volume energy scale $R^{-1}$ and the microscopic cutoff $\Delta_{\rm LL}$, i.e. describes going to long wavelengths. Calculations at fixed $R$ with $Q\to\infty$ describe a different choice of units or a different trajectory unless the dimensionful parameters are rescaled accordingly.

\subsection{The two integration orders}\label{sub:twoorders}
For real neutral variables $X^I$, with $I$ labeling interaction channels, consider a Euclidean action:
\begin{equation}\label{eq:mothergeneral}
	S_{\rm M}=\int\dd\tau\,\dd A\, \bar\Psi\bigl(\slashed D_A+\mathsf M_0+g_I X^I T_I\bigr)\Psi +\frac12 X^I(\mathcal K_0^{-1})_{II'}X^{I'}+S_{\rm ng}[X].
\end{equation}
The kernel includes its spacetime integrals, and $S_{\rm ng}[X]$ accounts for possible non-Gaussianities. It is also understood that $J_I=\bar\Psi T_I\Psi = \bar\Psi_a (T_I)_{ab}\Psi_b$, where $a,b$ label flavors. A spatially local but nondynamical HS field is permitted: microscopic Lorentz symmetry is not required for a relativistic IR limit. The charged Gaussian integral gives
\begin{equation}\label{eq:detgeneral}
	\Gamma[X]=S_X-\operatorname{Tr}\log\bigl(\slashed D_A+\mathsf M(X)\bigr).
\end{equation}
The trace-log is the Gaussian charged determinant with the specified occupied state prescription (which is often enforced automatically in the zero-temperature system). Below we use its zero-temperature vacuum at the stated filling. This determinant is exact at fixed $X$ when the charged action is quadratic. Its derivative expansion determines the $X$ EFT; the remaining functional integral over $X$ captures the conventional RG flow to the CFT point. Additional charged interactions (such as the density improvement channel discussed in the Ising case) must either be kept through their own HS fields or treated with a separately justified expansion.

If $S_{\rm ng}=0$, the opposite integral is the identity
\begin{equation}\label{eq:HSidentity}
	\int\mathcal DX\,\exp\left[-\frac12X\mathcal K_0^{-1}X-gXJ\right] =\mathcal N\exp\left[\frac{g^2}{2}J\mathcal K_0J\right].
\end{equation}
A positive real covariance therefore gives an attractive Hamiltonian channel $-g^2J\mathcal K_0J/2$. A repulsive channel $+J VJ/2$ can instead be decoupled by $\frac12\rho V^{-1}\rho+\ii\rho J$. This is an auxiliary variable, not a physical scalar with an imaginary Yukawa coupling, and we tend to keep it massive so that it does not enter the IR physics.

At finite $N$, any specified two-body interaction on the LLL can be decomposed into such channels, furnishing an exact band-limited HS representation, i.e. giving the ``reverse route''. Its inverse kernel ${\mathcal K}_0^{-1}$ can be nonlocal, thus an exact HS rewriting alone does not provide a local mother theory (which is fine). Conversely, a local kernel $\mathcal K_0^{-1}=r_0-\kappa\nabla^2$ would lead to a derivative expansion for ${\mathcal K}_0$, generating higher pseudopotentials in $\cH_{\rm fuzzy}$. Therefore, in the ``reverse route'', we must choose whether we match the two-body interaction in $\cH_{\rm fuzzy}$ with the published finite-pseudopotential Hamiltonians, or prefer to keep locality of ${\mathcal K}_0^{-1}$ while allowing higher pseudopotentials (we usually choose the former, writing a local ${\mathcal K}_0=g_1 + g_2 \boldsymbol\nabla^2$ and allowing the nonlocal Gaussian kernel ${\mathcal K}_0^{-1}$ in the ``reverse route'' mother action).

We define the pseudopotentials \cite{Haldane1983} through the two-particle matrix elements,
\begin{equation}\label{eq:pseudomatch}
	V_J^{ab;cd}=\langle(ss)JM;ab|H_{\rm int}|(ss)JM;cd\rangle, \qquad r=2s-J,
\end{equation}
which describe an interaction of two particles of flavor $(c,d)$ into the two particles of flavor $(a,b)$. Here $J$ is total angular momentum of a pair and $M$ -- its projection. Rotational invariance guarantees that $V_J^{ab;cd}$ is independent of $M$. When the interaction acts as the identity in flavor space, this reduces to the usual scalar pseudopotentials, which are more often labeled by the relative angular momentum $r=2s-J$ (rather than total $J$) as $V_r$, and that is the convention taken in most of the literature, where $V_0$ and $V_1$ refer to $r=0$ and $r=1$. Also, the appropriate bosonic or fermionic exchange projection must be imposed (for fermions, even $r$ requires an antisymmetric flavor state, and odd $r$ -- symmetric). We use empty-vacuum normal ordering, and for a static scalar covariance, the projected interaction is explicitly:
\begin{align}
	H_{\rm int}&=-\frac{g^2}{2}
	\sum_{\substack{m_1m_2m_3m_4\\abcd, II'}}
	{\cal V}^{II'}_{m_1m_2;m_3m_4}\,T_{I,ab}T_{I',cd}\,
	c^\dagger_{m_1a}c^\dagger_{m_2c}c_{m_3d}c_{m_4b},
	\label{eq:explicitprojectedH}\\
	{\cal V}^{II'}_{m_1m_2;m_3m_4}
	&=\int\dd A_x\dd A_y\,K^{II'}(x,y)
	u_{m_1}^*(x)u_{m_2}^*(y)u_{m_3}(y)u_{m_4}(x).
	\label{eq:orbitalinteraction}
\end{align}
Multiple mediator channels are summed with their own flavor matrices and covariances. This formula, or its Clebsch--Gordan transform in Eq.~\eqref{eq:pseudomatch}, is the finite-dimensional Hamiltonian that can be used in the ``reverse route.''

Normal ordering can also change a one-body term. With $\mathcal P(x,y)=\sum_m u_m(x)u_m^*(y)$, the contraction is
\begin{equation}\label{eq:normalordering}
	J_I(x)J_{I'}(y)=:J_I(x)J_{I'}(y): +\mathcal P(x,y)\psi^\dagger(x)T_I T_{I'}\psi(y),
\end{equation}
where we already consider LLL-projected fermions:
\begin{equation}
	J_I(x) = \psi^\dagger_a(x) (T_I)_{ab}\psi_b(x),\qquad \psi_a(x)=\sum_m u_m(x) c_{ma}, \qquad \{\psi_a(x), \psi_b^\dagger(y)\}=\delta_{ab}{\mathcal P}(x,y).
\end{equation}
The one-body term is an inconsequential constant at fixed total filling only when the summed one-body operator is proportional to total number. Otherwise it must be kept in the analyses.

These expressions also lead to an exact HS transformation at finite $N$. Keep only the harmonics $0\le L\le N-1$ of an instantaneous Gaussian mediator $X$, and couple its projected multiplication operator $\widehat X = P_{\LLL} X P_\LLL$ (which is how $X$ acts on the LLL) to the LLL fermions. For the real Ising channel, for example, integrating out LLL fermions gives:
\begin{equation}\label{eq:exactbanddet}
	\Gamma_N[X]=\frac12\widehat XK^{-1}\widehat X -\Tr_{V_s\otimes\mathbb C^2\otimes\tau} \log\!\left[\partial_\tau+m\sigma^x+g\widehat X\sigma^z\right] +\mathrm{const}.
\end{equation}
Here $K$ is the chosen finite-$N$ covariance (usually $g_1 + g_2 \boldsymbol{\nabla}^2$), with positive eigenvalues for a real attractive channel. Eliminating $X$ instead gives Eq.~\eqref{eq:explicitprojectedH} exactly. Equation~\eqref{eq:exactbanddet} has finite spatial cutoff $N-1$ but remains a function of continuous time; its local low energy expansion leads to an EFT of $X$. HLLs are absent from this definition, so it has no HLL-elimination error. A local relativistic mother, on the contrary, contains HLLs and must justify their removal as a separate controlled approximation (justified by the large gap $\Delta_{\rm LL}$).

Finally, for non-Gaussian mediators the integration gives:
\begin{equation}\label{eq:cumulants}
	S_{\rm ind}[J]=-\log\left\langle e^{-g\int XJ}\right\rangle_X =-\sum_{p\geq1}\frac{(-g)^p}{p!}\int C_X^{(p)}(1,\ldots,p)J_1\cdots J_p.
\end{equation}
There is no general reason for a two-body truncation to be valid, so if needed, it should be justified on a case by case basis.

\subsection{Berry curvature in the magnetic background}\label{sub:berrytest}
Let $P(X)$ project onto the occupied flavor subspace of the LLL. Choose a local orthonormal occupied frame $U$, with $P=UU^\dagger$. The adiabatic response \cite{Berry1984} of the occupied state is:
\begin{equation}\label{eq:Berryprojector}
	S_B=\rho_0\int\dd\tau\,\dd A\,\operatorname{tr}(U^\dagger\partial_\tau U), \qquad \mathcal F=\ii\operatorname{tr}\bigl(P\,\dd P\wedge\dd P\bigr).
\end{equation}
The former is Berry connection; the latter is Berry curvature. The occupied state, being non-degenerate, has abelian Berry connection, hence the trace. The LLL contribution is often the decisive one, with the HLL contributions canceling (due to the HLLs being non-chiral); this cancellation, however, must be checked for arbitrary masses and fillings.

A real rank-one occupied eigenvector gives $U^T\dd U=0$ locally. For higher rank, a real occupied projector still has vanishing traced Berry curvature, since for symmetric $P$, $\dd P\wedge \dd P$ is an antisymmetric matrix, so ${\rm tr}(P\,\dd P\wedge \dd P)=0$. A quaternionic occupied bundle also has zero trace curvature: its structure group is symplectic and its determinant line is trivial. These facts explain, respectively, the reference flavor vector/rotor and symplectic constructions below. Notably, they do not exclude Wess--Zumino terms.

For a single real field, $f(\phi)\partial_\tau\phi$ is a total derivative in a smooth patch. For a complex field, the independent term
\begin{equation}\label{eq:complexberry}
	\frac b2\bigl(\Phi^*\partial_\tau\Phi-\Phi\partial_\tau\Phi^*\bigr)
\end{equation}
is not a total derivative and may be generated. For example, it obstructs a would-be $O(2)$ Wilson--Fisher point of the easy-plane doublet in \cite{MotherPRL}. A $U(1)$ or $Z_3$ symmetry permits it; an internal charge reflection $\Phi\mapsto\Phi^*$ forbids it. That reflection must act within the \emph{same flux and filling sector} to be the symmetry of background.

In our analyses the Berry term is a simple diagnostic tool of the IR limit. The Berry term should vanish in order for the Lorentz symmetry to have a chance of being restored in the IR.

\subsection{Cylinder matching and the operator dictionary}

For a free canonically normalized noncompact scalar, the improved cylinder action~\cite{CCJ} contains
\begin{equation}\label{eq:conformalmass}
	\frac12(\partial\varphi)^2+\frac12\xi\mathcal R\varphi^2, \qquad \xi=\frac18,\quad \mathcal R=\frac2{R^2},\quad E_L=\frac vR\left(L+\frac12\right).
\end{equation}
The curvature couplings are not universal and are different in interacting theories. There one tunes the relevant couplings in the curved-background EFT and tests the cylinder spectrum $E_a-E_0=(v/R)\Delta_a$. The background flux renormalizes scaling fields and the velocity even when all CFT operators are electromagnetically neutral.

When $X$ is Gaussian, its integration also fixes the operator map, including contact terms:
\begin{equation}\label{eq:dictionary}
	X=-g\mathcal K_0J+\eta,\qquad \langle X_1X_2\rangle_c=g^2(\mathcal K_0)_1(\mathcal K_0)_2 \langle J_1J_2\rangle_c+\mathcal K_0(1,2),
\end{equation}
where $\eta$ is Gaussian noise with covariance ${\mathcal K}_0$. When the latter is a differential contact covariance (such as our $g_1 + g_2\boldsymbol{\nabla}^2$), the last term is contact term, and the first mixes $J$ with its derivatives. For a finite range covariance both kernel factors and the Gaussian term must be retained to obtain the correct correlators. Finally, an odd mediator obeys analogous Grassmann identity with an odd, appropriately charged source and odd Gaussian noise.

\section{Real scalar mother theories}\label{sec:real}

\subsection{The Ising mother, its symmetries, and the projected Hamiltonian}\label{sub:isingmother}
The Ising mother theory from \cite{MotherPRL} contains two charge $+1$ Dirac fermions, $\Psi=(\Psi_1,\Psi_2)^T$, and a neutral real scalar $\phi$. Pauli matrices act on the two flavors, and the minimal action is:
\begin{equation}\label{eq:IsingMother}
	\cL_I=\bar\Psi(\slashed D_A + m\sigma_F^x+g\phi\sigma_F^z)\Psi + \frac12\phi\mathcal K_0^{-1}\phi,\qquad \mathcal K_0^{-1} =r_0-\kappa_s\boldsymbol\nabla^2.
\end{equation}
Here we choose an instantaneous kernel. Its covariance must be positive
on the retained harmonics if $\phi$ has a real Gaussian measure. We take
$m>0$; the opposite sign is related by a flavor rotation.

Besides spatial rotations and the conserved electric charge, the action
has the Ising symmetry
\begin{equation}\label{eq:IsingSymmetry}
	\phi\mapsto-\phi,\qquad \Psi\mapsto\sigma^x_F\Psi.
\end{equation}
The discrete spacetime symmetries follow the conventions of \cite{MotherPRL}. Charge conjugation $C$ reverses electric charge,
$P$ reverses the spatial orientation, and $T$ is time reversal.
The three-dimensional Dirac mass terms are odd under $P$ and $T$.
Conjugation by $\ii\sigma^y_F$ reverses both $\sigma^x_F$ and $\sigma^z_F$, so at zero background field the mass terms preserve
$\widetilde P=(\ii\sigma^y_F)P$ and $\widetilde T=(\ii\sigma^y_F)T$.
Each of $C$, $\widetilde P$, and $\widetilde T$ reverses the magnetic flux. Their products
\begin{equation}\label{eq:IsingFluxSymmetries}
	\widetilde{CP}=C(\ii\sigma^y_F)P,\qquad \widetilde{CT}=C(\ii\sigma^y_F)T
\end{equation}
preserve the chosen flux sector and act on $\phi$ as ordinary spatial
parity and time reversal. After LLL projection they give, respectively,
the orientation reversing unitary and orientation preserving antiunitary
particle--hole symmetries of the fuzzy Ising Hamiltonian~\cite{ZhuIsing,MotherPRL}.

The original fuzzy Ising model also contains a repulsive interaction in
the total density~\cite{ZhuIsing} that improves convergence to the CFT point. To include it, introduce an independent real auxiliary $\rho$:
\begin{equation}\label{eq:densityMother}
	\cL_{I,\rho}=\cL_I+\frac12\rho K_\rho^{-1}\rho+\ii g_\rho\rho\,\bar\Psi\Psi.
\end{equation}
The quadratic form is positive, and the integral over $\rho$ is along
the real contour; the imaginary coupling gives the repulsive sign.
A background subtraction in the singlet bilinear (i.e., normal ordering of $\bar\Psi\Psi$ in Dirac vacuum) is understood so that
its LLL projection is $\delta n^0=n^0-\rho_0$. For a rotationally invariant kernel and fixed total particle number, using $n^0$ instead only shifts Hamiltonian by a constant. In LLL-projected expressions we write $n^a=\psi^\dagger\sigma^a_F\psi$, $n^0=\psi^\dagger\psi$, and
$N^x=\int\dd A\,n^x$. Eliminating both auxiliaries gives
\begin{align}\label{eq:IsingH}
	H_I&=mN^x-\frac{g^2}{2}\int\dd A_x\dd A_y\,:n^z(x)\mathcal K_0(x,y)n^z(y):+H_\rho,\cr
	H_\rho&=\frac12\int\dd A_x\dd A_y\,:\delta n^0(x)V_\rho(x,y)\delta n^0(y):,\qquad V_\rho=g_\rho^2K_\rho.
\end{align}
We call $H_\rho$ the density channel. In the normalization of Eq.~\eqref{eq:IsingH}, the equal-kernel
choice $V_\rho=g^2\mathcal K_0$ reproduces the relative interaction strengths of Ref.~\cite{ZhuIsing}. Transverse field is $h=-m$.

For an exact reverse construction, choose the two-pseudopotential approximation from \cite{ZhuIsing}, namely, take covariance $\cK_0=g_1 + g_2 \boldsymbol{\nabla}^2$ as described in Section \ref{sub:twoorders}.\footnote{The derivative interaction should not be discarded. The strictly contact Hamiltonian has an $SU(2)_F$ invariant interaction. For positive contact strength, $g^2 g_1>0$, its half-filled ground state is fully polarized along the transverse field for $m>0$, and its neutral gap is exactly $2m$. The case of $m<0$ is analogous. It therefore has no Ising transition, and it is crucial to keep nonzero $g_2$ to get a transition.}
A local inverse kernel $\cK_0^{-1}=r_0-\kappa_s\boldsymbol\nabla^2$ instead gives a finite range
potential with higher pseudopotentials. The two choices
have the same symmetry but need not have the same critical couplings.

\subsection{Integrating out the fermions at fixed flux}\label{sub:isingflux}
We first set $g_\rho=0$; the density interaction is discussed separately
below. In a constant $\phi$ background, the flavor masses are $\pm M(\phi)$, where $M(\phi)=\sqrt{m^2+g^2\phi^2}$. The negative LLL flavor is filled, giving $N_e=N$. The gap also survives a nonuniform static $\phi$. With Dirac Hamiltonian matrices $\boldsymbol\alpha,\beta$, write
$H=H_0[\phi]+m\beta\sigma^x_F$ and $H_0[\phi]=\boldsymbol\alpha\cdot(-\ii\boldsymbol D)+g\beta\phi\sigma^z_F$. Since $\{H_0,\beta\sigma^x_F\}=0$, one has $H^2=H_0^2+m^2$, and hence $|E|\ge m$. The derivative expansion is
therefore performed below the charged gap $\Delta_c=m$, where we expect an EFT of the form:
\begin{equation}\label{eq:IsingFluxEFT}
	\cL_{\phi}=\frac{Z_\tau}{2}(\partial_\tau\phi)^2+\frac{Z_s}{2}(\boldsymbol\nabla\phi)^2+\frac{r_2}{2}\phi^2+\frac{u_4}{4!}\phi^4+\cdots.
\end{equation}
For comparison, the flat space determinant gives the exact effective potential:
\begin{equation}\label{eq:flatpot}
	\Delta v_f^{\rm flat}=\frac{(m^2+g^2\phi^2)^{3/2}-m^3}{3\pi},
\end{equation}
up to a quadratic counterterm. Its quartic and (separately computed, assuming bare values $Z_\tau^{(0)}=Z_s^{(0)}=\kappa_0$) kinetic coefficients are:
\begin{equation}\label{eq:flatcoeff}
	u_{4,f}^{\rm flat}=\frac{3g^4}{\pi m}>0,\qquad Z_f^{\rm flat}=\kappa_0+\frac{g^2}{4\pi m}.
\end{equation}
At fixed magnetic flux the coefficients must instead be computed from the spherical Landau levels.
We quote the determinant and response derived in the Supplemental
Material of Ref.~\cite{MotherPRL}; Landau level sums of this kind also
appear in the magnetic effective actions of Ref.~\cite{GusyninMiranskyShovkovy1995}.
The uniform ground state energy shift is:
\begin{equation}\label{eq:fluxpotential}
	\Delta\Omega_f(\phi)=-N[M(\phi)-m]-2\sum_{n\ge1}d_n\left[\sqrt{\lambda_n^2+m^2+g^2\phi^2}-\sqrt{\lambda_n^2+m^2}\right]+\Delta\Omega_{\rm ct}.
\end{equation}
Define
\begin{equation}\label{eq:Sp}
	S_p=\frac N{m^p}+2\sum_{n\ge1}\frac{d_n}{(\lambda_n^2+m^2)^{p/2}}.
\end{equation}
The sums converge for $p>2$; $S_1$ requires the scalar mass counterterm.
Expansion per unit area gives:
\begin{equation}\label{eq:fluxquartic}
	r_{2,f}=-\frac{g^2}{\mathcal A}S_1\big|_{\rm ren},\qquad u_{4,f}=\frac{3g^4}{\mathcal A}S_3>0,\qquad u_{6,f}=-\frac{45g^6}{\mathcal A}S_5.
\end{equation}
In particular, magnetic flux does not reverse the sign of the induced
quartic. The negative sextic is just a coefficient in the small-field
expansion, the full potential is still bounded from below.

The full LLL quadratic response follows from the spectral sum and orbital
trace in Appendix~\ref{app:spectral}:
\begin{equation}\label{eq:LLLresponse}
	\Pi_0(\ii\omega,L)=-\frac{4\rho_0g^2m}{\omega^2+4m^2}\,p_L(N),\qquad p_L(N)=\prod_{j=1}^{L}\frac{N-j}{N+j},\qquad p_0=1.
\end{equation}
Its frequency dependence is the response of an occupied flavor to the
empty flavor, separated by energy $2m$. The factor $p_L$ is the squared
LLL matrix element of the spherical harmonic. Expanding
$p_L=1-L(L+1)/N+O(N^{-2})$ at fixed $L$ yields:
\begin{equation}\label{eq:LLLkinetic}
	Z_{\tau,0}=\frac{\rho_0g^2}{4m^3},\qquad Z_{s,0}=\frac{\rho_0g^2R^2}{mN}=\frac{g^2}{4\pi m}.
\end{equation}
Both coefficients are positive, showing, for example, that in the reverse route with no bare kinetic term nor bare stiffness, a healthy $\phi$ EFT is generated just by integrating out LLL.

At $L=0$ the HLL contribution is
\begin{equation}\label{eq:HLLresponse}
	\Pi_H(\ii\omega,0)=-\frac{8g^2}{\mathcal A}\sum_{n\ge1}\frac{d_nE_n}{\omega^2+4E_n^2},\qquad Z_{\tau,H}=\frac{g^2}{4\mathcal A}S_3^H>0,
\end{equation}
where $E_n=\sqrt{m^2+\lambda_n^2}$ and $S_3^H$ excludes the LLL.
Spatial HLL corrections require the nonuniform matrix elements; the
magnetic background does not permit inferring them from Lorentz symmetry.

The occupied flavor eigenvector of $m\sigma^x+g\phi\sigma^z$ can be chosen
real and smooth, so its Berry connection vanishes. More generally, a
term $f(\phi)\partial_\tau\phi$ for one real field is a total derivative.
The resulting local action, therefore, indeed has the Ising form \eqref{eq:IsingFluxEFT}.
It can flow to the Ising Wilson--Fisher fixed point~\cite{WilsonFisher}
when the relevant even coupling is tuned and the higher terms do not
produce a competing transition.

\subsection{What the density interaction changes}\label{sub:isingdensity}
Equation~\eqref{eq:densityMother} characterizes the density channel. Provided $\rho$ remains massive, eliminating it together with the charged fields changes the coefficients of \eqref{eq:IsingFluxEFT} without adding another light field. As $\phi$ is tuned to criticality, the field $\rho$ turns out to remain around mass scale $\Delta_c$. Thus it cannot be treated reliably by first integrating out fermions, truncating the $\phi, \rho$ EFT at low frequency and then integrating that truncated action over all frequencies. The contribution of $\rho$ must be computed with the charged response at the same scale, which may be hard analytically but is doable numerically \cite{MotherPRL}.

For equal spatial kernels, the relative density strength $\alpha$ has the elementary decomposition:
\begin{align}\label{eq:alphaChannels}
	\alpha n^0(x)n^0(y)-n^z(x)n^z(y)={}&(\alpha-1)[n_\uparrow(x)n_\uparrow(y)+n_\downarrow(x)n_\downarrow(y)]\cr
	&+(\alpha+1)[n_\uparrow(x)n_\downarrow(y)+n_\downarrow(x)n_\uparrow(y)].
\end{align}
Thus $\alpha=1$ removes interactions between particles of the same flavor (intralayer). The conformal perturbation theory analysis in \cite{MotherPRL}, using the method from \cite{LauchliCPT}, finds that along the tuned critical line, setting $\alpha=1$ suppresses the leading irrelevant scalar coupling by an order of magnitude. This is a numerical result that does not follow from the flavor identity above.

\subsection{The tricritical Ising model}\label{sec:tricritical}
Reference \cite{TaylorTricritical} has an intralayer potential $U$ and the interlayer potential $V$. In our conventions, with spatial integrals understood, it is:
\begin{align}\label{eq:tricriticalChannels}
	H_{\rm tri}&=-hN^x+2n_\uparrow Vn_\downarrow+\lambda\sum_{a=\uparrow,\downarrow}:n_aUn_a:\cr
	&=-hN^x+\frac12:n^0(V+\lambda U)n^0:-\frac12:n^z(V-\lambda U)n^z:.
\end{align}
Its mother theory is precisely \eqref{eq:densityMother}, with unequal Ising and density kernels:
\begin{equation}\label{eq:tricriticalKernels}
	g^2\mathcal K_0=V-\lambda U,\qquad g_\rho^2K_\rho=V+\lambda U,\qquad m=-h.
\end{equation}
These specify the full projected interaction, including the higher pseudopotentials used in \cite{TaylorTricritical}.
%If a kernel has both signs, diagonalize it on the finite set of harmonics and decouple each eigenvalue using a real auxiliary with a real or imaginary coupling, as appropriate. The Gaussian quadratic forms can all be taken positive. 
The Hamiltonian preserves the Ising and particle--hole symmetries described above. The infrared EFT, after $\Psi$ and $\rho$ are eliminated, can now be tuned to the tricritical regime:
\begin{equation}\label{eq:tricritical}
	\cL_{\rm tri,IR}=\frac12(\partial\phi_c)^2+\frac{r_c}{2}\phi_c^2+\frac{\lambda_4}{4!}\phi_c^4+\frac{\lambda_6}{6!}\phi_c^6+\cdots,\qquad r_c=\lambda_4=0,\quad\lambda_6>0.
\end{equation}
On the sphere, $r_c$ must be tuned to the conformal value $\frac{1}{4R^2}$, and this EFT flows to the Gaussian scalar \cite{TaylorTricritical} as the positive sextic is marginally irrelevant in 3D. There is still only one real scalar, so no Berry term obstructs this limit. The two independent kernels allow the relevant couplings to be varied independently (unlike at $\cK_0=\cK_\rho$). The critical values and the positive sextic are not computed analytically here. The evidence that the specified two-body Hamiltonian reaches the tricritical regime is the phase diagram and conformal spectrum computed in \cite{TaylorTricritical}.

\subsection{Lee--Yang continuation}\label{sec:leeyang}
The fuzzy Lee--Yang models add an imaginary longitudinal field to the Ising Hamiltonian \cite{FanLeeYang,ArguelloLeeYang,EliasLeeYang}. Their mother action is \eqref{eq:densityMother}, with a source $\ii\lambda\phi$ added. The Gaussian integral over $\phi$ gives:
\begin{equation}\label{eq:LYreverse}
	-\frac12(g n^z+\ii\lambda)\mathcal K_0(g n^z+\ii\lambda)=-\frac{g^2}{2}n^z\mathcal K_0n^z-\ii g\lambda\,n^z\mathcal K_0 1+\mathrm{const}.
\end{equation}
Since $\mathcal K_0 1=K_{L=0}$, the source produces exactly an imaginary longitudinal field. The separate density channel is unchanged. The source breaks the Ising $\Z_2$, but preserves its product with the antiunitary symmetry that complex conjugates $\ii$.

In the scalar EFT, approach the Lee--Yang edge from the disordered Ising phase and expand around the complex saddle. A field shift removes the quadratic term, leaving the conventional edge action:
\begin{equation}\label{eq:LYEFT}
	\cL_{\rm LY}=\frac12(\partial\varphi)^2+\ii t\varphi+\frac{\ii g_3}{3!}\varphi^3+\cdots.
\end{equation}
Tuning $t$ gives the Lee--Yang critical theory; the cited fuzzy sphere studies test this identification through its spectrum. A single scalar again has no Berry term. The charged determinant must remain gapped near the saddle: if $m^2+g^2\phi_*^2=0$, the gap closes and this derivation no longer applies. The continuation is therefore made within the scalar EFT below the charged threshold.

\section{Easy-plane mothers, saturation, and free scalars}\label{sec:easyplane}

\subsection{The doublet, its symmetries, and the Berry term}
The easy-plane mother introduced in \cite{MotherPRL} replaces the real Ising mediator by a complex field $\Phi$. It contains a charge-one Dirac flavor doublet $\Psi=(\Psi_1,\Psi_2)^T$, while $\Phi$ remains electrically neutral:
\begin{equation}\label{eq:EPmother}
	\cL_\ep=\bar\Psi(\slashed D_A+\mu\sigma^z)\Psi+g(\Phi n^-+\Phi^*n^+)+\Phi^*K_\Phi^{-1}\Phi,\qquad n^\pm=\bar\Psi\sigma^\pm\Psi,\qquad \sigma^\pm=\frac{\sigma^x\pm\ii\sigma^y}{2}.
\end{equation}
The Pauli matrices act on flavor. Besides spatial rotations and the charged $U(1)_V$ that couples to the background flux, the theory preserves:
\begin{equation}\label{eq:EPflavor}
	U(1)_F:\qquad\Psi\mapsto e^{\ii\alpha\sigma^z/2}\Psi,\qquad\Phi\mapsto e^{-\ii\alpha}\Phi.
\end{equation}
The reflection $R_F:\Psi_1\leftrightarrow\Psi_2$, $\Phi\leftrightarrow\Phi^*$ sends $\mu\mapsto-\mu$. At $\mu=0$ it extends $U(1)_F$ to $O(2)_F$; at nonzero $\mu$ it relates two members of the family. With $C,P,T$ defined as in Sec.~\ref{sub:isingmother} and Ref.~\cite{MotherPRL}, the zero-flux theory preserves charge conjugation $C$ and the combinations $\widetilde P=R_FP$, $\widetilde T=R_FT$, although $P$ and $T$ separately reverse the mass. The magnetic background preserves only $\widetilde{CP}=C\widetilde P$ and $\widetilde{CT}=C\widetilde T$ among these discrete transformations. Unlike an internal reflection $R_F$ preserved at zero $\mu$, the $\widetilde{CP}$ and $\widetilde{CT}$ symmetries permit the Berry term below.

For $\mu\ne0$, the LLL mass matrix $h_\ep=\left(\begin{smallmatrix}\mu&g\Phi^*\\g\Phi&-\mu\end{smallmatrix}\right)$ has one negative eigenvalue per orbital. Filling these $N$ states gives a charged gap $|\mu|$ at $\Phi=0$. The static determinant is obtained from Eq.~\eqref{eq:fluxpotential} by replacing $m\mapsto|\mu|$ and $\phi^2\mapsto|\Phi|^2$. Its positive quartic would suggest an $O(2)$ model-like complex scalar EFT, but its kinetic term must also be computed. The exact uniform LLL response, derived in the Supplement of \cite{MotherPRL}, is:
\begin{equation}\label{eq:EPberry}
	\Pi_{\Phi,\LLL}(\ii\omega,0)=-\frac{\rho_0g^2}{2|\mu|+\ii\operatorname{sgn}(\mu)\omega},\qquad \cL_B=c(\Phi^*\partial_\tau\Phi-\Phi\partial_\tau\Phi^*),\qquad c=-\operatorname{sgn}(\mu)\frac{\rho_0g^2}{8\mu^2}.
\end{equation}
The paired HLLs give an even function of frequency and do not change $c$. For $\Phi=(\pi_1+\ii\pi_2)/\sqrt2$, the Berry term is $\ii c(\pi_1\partial_\tau\pi_2-\pi_2\partial_\tau\pi_1)$, so it is imaginary in Euclidean signature. Its covariant origin is a topological term $-\frac{\operatorname{sgn}(\mu)g^2}{8\mu^2} \frac{F}{2\pi}\wedge(\bar\Phi\dd\Phi-\Phi\dd\bar\Phi)$ \cite{MotherPRL,Dedushenko:2022pem,Kapustin:2020eby}. Note that a single occupied state Berry connection \cite{Berry1984} is $U^\dagger\dd U=-\frac{\operatorname{sgn}(\mu)g^2(\bar\Phi\dd\Phi - \Phi\dd\bar\Phi)}{4(\mu^2 + g^2|\Phi|^2 + \mu\sqrt{\mu^2 + g^2|\Phi|^2})}$, so up to non-Gaussian corrections, our topological term describes coupling of an occupied state Berry connection to $F/(2\pi)$.

To find the excitation energies, retain both time derivatives in the quadratic action,
\begin{equation}\label{eq:EPquadratic}
	\cL_\Phi^{(2)}=Z_t|\partial_\tau\Phi|^2+Z_s|\nabla\Phi|^2+r|\Phi|^2+\cL_B,\qquad Z_t,Z_s>0.
\end{equation}
After continuation to real frequency, the determinant of the inverse propagator vanishes when:
\begin{equation}\label{eq:EPpoleequation}
	[Z_sk^2+r-Z_tE^2]^2-(2cE)^2=0.
\end{equation}
The two circular polarizations (modes with opposite $U(1)_F$ charges) therefore obey equations $Z_tE^2\pm2cE-(Z_sk^2+r)=0$. For $r>0$, in the uncondensed phase, their positive roots are:
\begin{equation}\label{eq:EPbranches}
	E_\pm(k)=\frac{\sqrt{(2c)^2+4Z_t(Z_sk^2+r)}\pm2|c|}{2Z_t},\qquad E_-(k)\big|_{r=0}=\frac{Z_s}{2|c|}k^2+O(k^4).
\end{equation}
Thus the gapless mode has $z=2$. The upper root is the opposite circular polarization of this quadratic truncation. It represents a physical excitation only if it lies within the EFT's frequency range. For the purely LLL-induced coefficients, $Z_t=\rho_0g^2/(8|\mu|^3)$ gives $E_+(0)=2|\mu|$ at $r=0$, precisely the scale at which the expansion of Eq.~\eqref{eq:EPberry} fails. The quadratic truncation does not establish an additional pole there. Nor can one set $\mu=0$ to remove $c$: that closes the charged gap and makes the static determinant nonanalytic at $\Phi=0$.

\subsection{The projected Hamiltonian and exact saturation}
An instantaneous Gaussian $\Phi$ gives:
\begin{align}\label{eq:EPH}
	H_\ep&=-g^2:n^+K_\Phi n^-:-h_zN^z\nonumber\\
	&=-\frac12\int\dd A_x\dd A_y\,U_\perp(x,y):[n^x(x)n^x(y)+n^y(x)n^y(y)]:-h_zN^z.
\end{align}
The parameters are $U_\perp=g^2K_\Phi/2$ and $h_z=-\mu$. Throughout this section, the densities are LLL projected and $N^z=\int\dd A\,n^z$. We take precisely the two pseudopotentials $V_0,V_1\ge0$ of Ref.~\cite{MotherPRL}, at total particle number $N$. A density interaction or additional pseudopotentials would require a separate saturation analysis.

We summarize the one-magnon solution and the proof for all magnon numbers from the Supplement of \cite{MotherPRL}. The orbital recoupling in the one-magnon calculation is also familiar from quantum Hall spin waves~\cite{NakajimaAoki1994}. The fully $z$-polarized (i.e., zero-magnon) state $|F\rangle$ has $E_F=-Nh_z$. For large enough $h_z$, it is the ground state. The one-magnon states are separated by gaps $2(h_z-h_1)+\varepsilon_L$, where $h_1$ is the saturation field, and $\varepsilon_L$ describes the one-magnon dispersion. Writing $p_L=\prod_{j=1}^L(N-j)/(N+j)$, we have:
\begin{align}\label{eq:magnondisp}
	h_1&=V_1\left(4-\frac6N\right),\nonumber\\
	\varepsilon_L&=\frac2N\left\{(2N-1)V_0(1-p_L)+(2N-3)V_1\left[1+p_L\left(\frac{2L(L+1)}{N-1}-1\right)\right]\right\}.
\end{align}
In particular, $\varepsilon_0=0$, so the uniform ($L=0$) one-magnon gap closes at the saturation field $h_z=h_1$. To show that no higher-magnon state destabilizes the vacuum at a larger field, introduce the pair annihilators and spin-raising multipoles:
\begin{align}\label{eq:EPpairOperators}
	A_{rq}&=\sum_{m m'}C^{J_rq}_{sm,sm'}c_{m'\uparrow}c_{m\downarrow},\qquad \mathcal C_r=\sum_qA_{rq}^\dagger A_{rq},\qquad J_r=N-1-r,\nonumber\\
	B_{LM}&=\sum_{m m'}(-1)^{s-m'}C^{LM}_{sm,s,-m'}c^\dagger_{m'\uparrow}c_{m\downarrow}.
\end{align}
The interaction is $2V_0\mathcal C_0-2V_1\mathcal C_1$. Anticommuting a creation operator past an annihilator and recoupling the orbital coefficients expresses $\mathcal C_1$ in terms of $N_\downarrow$, $\mathcal C_0$, and $B_{LM}^\dagger B_{LM}$. At $h_z=h_1$, the term proportional to $N_\downarrow$ cancels, and we obtain:
\begin{equation}\label{eq:EPpositivity}
	H_\ep(h_1)-E_F(h_1)=2\left(V_0+\frac{2N-3}{2N-1}V_1\right)\mathcal C_0+\frac{2h_1}{N-1}\sum_{L\ge1,M}p_LL(L+1)B_{LM}^\dagger B_{LM}\ge0.
\end{equation}
Every term is non-negative in every particle-number sector: for example, $\langle\mathcal C_0\rangle=\sum_q\|A_{0q}|\psi\rangle\|^2$. The $k$-magnon gaps $\Delta_{k,\nu}$ acquire a further $2k(h_z-h_1)$ relative to the polarized state, so:
\begin{equation}\label{eq:EPallk}
	\Delta_{k,\nu}(h_z)\ge2k(h_z-h_1),\qquad h_{\rm sat}=h_1.
\end{equation}
The equality is reached by the uniform ($L=0$) one-magnon state. For $V_1>0$, the lowest higher-magnon gaps at $h_1$ satisfy the stronger finite-size result:
\begin{equation}\label{eq:EPfiniteGap}
	0<\delta_k(N)\le\frac{2h_1k(k-1)}{N-1},\qquad 2\le k\le N.
\end{equation}
The lower bound follows from Clebsch--Gordan completeness and the upper bound from taking average in the normalized state proportional to $(S^-)^k|F\rangle$; both are proved in \cite{MotherPRL}. At fixed $k$, the gaps $\delta_k(N)$ close as $N\to\infty$. Thus in the thermodynamic limit, the saturation transition at $h_z=h_1$ is consistent with a gas of non-interacting magnons. The flavor reflection $R_F$ maps $h_z$ to $-h_z$, so the opposite polarized state has saturation boundary $-h_1$. The positivity argument fixes both boundaries but does not exclude additional transitions between them. Thus the minimal phase diagram consistent with this analysis is:
\begin{center}
	\begin{tikzpicture}[x=0.9cm, font=\small]
		\draw[->] (-4,0) -- (4,0) node[right] {$h_z$};
		
		\fill (-1.4,0) circle (1.6pt) node[below=5pt] {$-h_1$};
		\fill ( 1.4,0) circle (1.6pt) node[below=5pt] {$h_1$};
		
		\node[above=7pt, align=center] at (-2.7,0)
		{Symmetric\\$-z$ polarized};
		\node[above=7pt, align=center] at (0,0)
		{SSB\\$U(1)_F$ broken};
		\node[above=7pt, align=center] at (2.7,0)
		{Symmetric\\$+z$ polarized};
	\end{tikzpicture}
\end{center}

\subsection{The planar \texorpdfstring{$z=2$}{z=2} limit}
At small $L/\sqrt N$, Eq.~\eqref{eq:magnondisp} gives
\begin{equation}\label{eq:EPmass}
	k_L^2=\frac{L(L+1)}{R^2},\qquad \varepsilon_L=\frac{k_L^2}{2M_*}+\cdots,\qquad M_*\longrightarrow\frac{N}{8R^2(V_0+3V_1)}.
\end{equation}
The last expression is the leading large-$N$ mass. In the planar (thermodynamic) limit of Sec.~\ref{sub:flux},
\begin{equation}\label{eq:EPplanar}
	N,R\longrightarrow\infty,\qquad B=\frac{N}{2R^2}\ \text{fixed},\qquad M_*\longrightarrow\frac{B}{4(V_0+3V_1)},\qquad k^2\ll2B.
\end{equation}
Both the magnon mass and magnetic scale stay finite. The familiar dilute-magnon EFT~\cite{BatyevBraginskii1984,SachdevSenthilShankar1994,NikuniOshikawaOosawaTanaka2000}:
\begin{equation}\label{eq:Schrodinger}
	\cL_{z=2}=b^*\left(\partial_\tau-\mu_b-\frac{\nabla^2}{2M_*}\right)b+\frac{u_R}{2}|b|^4+\cdots,\qquad \mu_b=2(h_1-h_z).
\end{equation}
On the repulsive scattering branch, the quartic is marginally irrelevant in two spatial dimensions, giving the Gaussian $z=2$ fixed point with logarithmic corrections \cite{FisherHohenberg1988,FisherWeichmanGrinsteinFisher1989,SachdevSenthilShankar1994,SachdevDunkel2006}. The exact bound \eqref{eq:EPpositivity} rules out a bound-state instability before saturation. The nonrelativistic state--operator correspondence uses a harmonic trap~\cite{NishidaSon,Boisvert:2025hex,Boisvert:2026mgw}; the spherical energies should not be interpreted as relativistic scaling dimensions.

\subsection{The ordered phase and conformal pinning}\label{sec:goldstone}
Choose $h_z=0$ in the easy-plane ordered phase and parameterize $\Phi=(v_0+\sigma)e^{\ii\theta}$, where $v_0$ is a fixed vev of $|\Phi|$. The Berry term
$\Phi^*\partial_\tau\Phi-\Phi\partial_\tau\Phi^*=2(v_0+\sigma)^2 \ii \partial_\tau\theta$ is a total time derivative $2\ii v_0\partial_\tau \theta$ for smooth configurations plus a coupling to the massive amplitude fluctuation $\sigma$. The latter generates a second-order time derivative when $\sigma$ is eliminated. Below its gap, the phase action is:
\begin{equation}\label{eq:phaseEFT}
	\cL_\theta=\frac{K_t}{2}(\partial_\tau\theta)^2+\frac{K_s}{2}(\nabla\theta)^2+\cdots,\qquad K_t,K_s>0.
\end{equation}
After $x_0=v\tau$, the canonical field is
\begin{equation}\label{eq:GoldstoneNormalization}
	v=\sqrt{K_s/K_t},\qquad \varphi=(K_tK_s)^{1/4}(\theta-\theta_0).
\end{equation}
A pinning potential of order $R^{-2}$ selects $\theta_0$. Tuning its quadratic term to the conformal mass $1/(4R^2)$ gives $E_L=(v/R)(L+\tfrac12)$, provided massive excitations and vortices lie above this spectral window. This construction gives a free relativistic scalar in the ordered phase~\cite{HeFreeScalar,MotherPRL}, despite the nearby $z=2$ saturation point being non-relativistic. The pinning potential is realized by adding $\frac{h_x}{R^2}n^x$ in the Hamiltonian, leading to the easy-plane version of the free scalar model:
\begin{equation}
	H_\ep = -\frac12 \int \dd A_x \dd A_y U_\perp(x,y):[n^x(x)n^x(y)+n^y(x)n^y(y)]: + \frac{h_x}{R^2} N^x.
\end{equation}
Assuming the two-pseudopotential $U_\perp(x,y)=[u_0+2U \boldsymbol{\nabla}^2]\delta(x,y)$, and using some LLL algebra, this can be written as \cite{MotherPRL}:
\begin{equation}
	H_\ep = \int \dd A \left[u_0 :(n^0)^2: - U :(n^x\boldsymbol\nabla^2 n^x+n^y\boldsymbol\nabla^2 n^y): + \frac{h_x}{R^2} n^x\right],\qquad u_0, U>0.
\end{equation}

\subsection{Relation to He's free scalar model}\label{sub:HeMother}

At filling $N_e=N$, He's Hamiltonian \cite{HeFreeScalar}, after adjusting to our conventions, is:
\begin{equation}\label{eq:HeHamiltonian}
	H_{\rm He}=\int\dd A\left[u_0:(n^0)^2:+U:n^z\boldsymbol\nabla^2n^z:+\frac{h_x}{R^2}n^x\right],\qquad u_0,U>0.
\end{equation}
The pinned easy-plane model \cite{MotherPRL} has the same contact and pinning terms, but derivative interaction
$-U\sum_{I=x,y}:n^I\nabla^2n^I:$. Thus:
\begin{equation}\label{eq:HeEPdifference}
	H_{\rm He}=H_{\ep}+U\mathcal D,\qquad
	\mathcal D=\int\dd A\sum_{I=x,y,z}:n^I\boldsymbol\nabla^2n^I:.
\end{equation}

Set first $h_x=0$ and consider the long-distance fluctuations in the ordered phase. Their infrared limit is the free Goldstone scalar of Section \ref{sec:goldstone}. The deformation $\mathcal{D}$ preserves $U(1)_F$ and the unbroken reflection, acting on the phase angle as $\theta\mapsto\theta+\epsilon_0$ and
$\theta\mapsto-\theta$, respectively. Shift symmetry forbids a phase potential. Apart from the identity, the only allowed operators of dimension at most three are kinetic terms. After canonical normalization, further
quadratic action terms have dimension at least five, and nonlinear derivative interactions have dimension at least six, modulo total derivatives and terms removable by field redefinitions. Consequently,
\begin{equation}\label{eq:HeIRDecomposition}
	\delta\cL_{\rm IR}=C_0+\frac{\delta K_t}{2}(\partial_\tau\theta)^2 +\frac{\delta K_s}{2}(\boldsymbol\nabla\theta)^2+\sum_{\Delta>3}c_\Delta\mathcal O_\Delta.
\end{equation}
Here $\mathcal{O}_\Delta$ has scaling dimension $\Delta$. The constant $C_0$ shifts the vacuum energy, while the kinetic terms change the velocity and field normalization. After the rescalings in \eqref{eq:GoldstoneNormalization}, the remainder is irrelevant. On the sphere, the conformal pinning must also be adjusted as in Section \ref{sec:goldstone}. This establishes irrelevance of $\mathcal{D}$ within the ordered phase after kinetic normalization. The numerical evidence for this statement was given in \cite{MotherPRL}, where the spectra of $H_{\rm ep}$ and $H_{\rm He}$ were shown to be extremely close.

\section{The $O(n)$ models, quantum rotors, and cubic anisotropy}\label{sec:on}
Now consider $n$ real scalar mediators $\phi_I$, $I=1\dots n$, playing role of $O(n)$ order parameter. We construct a mother theory coupling them to $n+1$ charged Dirac flavors and compute its response in magnetic flux. We then specify the interactions used in the literature to obtain $O(n)$, rotor, bilayer, cubic, and free scalar models~\cite{GuoON,DeyOtwo,DeyOthree,HanBilayer,StergiouCubic}.

\subsection{The singlet and \texorpdfstring{$n$}{n} charged flavors}

Let $\Psi=(\Psi_0,\Psi_1,\ldots,\Psi_n)^T$ be $n+1$ charge-one Dirac fields. The flavor $\Psi_0$ is an $O(n)$ singlet, while $(\Psi_1,\ldots,\Psi_n)$ transform in its defining representation. The following operators act only in this flavor space:
\begin{equation}\label{on:matrices}
	P_0=|0\rangle\langle0|,\qquad P_v=\sum_{I=1}^n|I\rangle\langle I|,\qquad T_I=|0\rangle\langle I|+|I\rangle\langle0|.
\end{equation}
Couple $n$ real neutral scalars $\bm\phi=(\phi_I)$, transforming in the same defining representation, to $\Psi$ through the usual Yukawa interactions, and also include a bare mass $m$ as follows:
\begin{equation}\label{on:mother}
	S_{O(n)}=\int\dd\tau\dd A\,\bar\Psi[\slashed D_A+m(P_v-P_0)+g\phi_IT_I]\Psi+\frac12\phi_IK_0^{-1}\phi_I,\qquad m>0.
\end{equation}
The repeated $I$ is summed over. This action has internal $O(n)$ symmetry: for $R\in O(n)$, $\Psi\mapsto\operatorname{diag}(1,R)\Psi$ and $\bm\phi\mapsto R\bm\phi$. An instantaneous choice $K_0^{-1}=r_0-\kappa_s\boldsymbol{\nabla}^2$ gives a local mother; a prescribed finite set of pseudopotentials instead fixes the ultralocal $K_{0}$ and makes its inverse nonlocal, as explained in Sec.~\ref{sub:twoorders}.

At $\bm\phi=0$, the singlet LLL has energy $-m$ and the other $n$ LLLs have energy $+m$. We fill the former, so $N_e=N=Q$. This naturally fixes the total particle number. At a uniform nonzero $\bm\phi$, only the singlet and the charged flavor parallel to $\bm\phi$ mix, giving the ``active'' mass matrix:
\begin{equation}\label{on:eigenvalues}
	\mathcal M_{\rm active}=\begin{pmatrix}-m&g|\bm\phi|\\g|\bm\phi|&m\end{pmatrix},\qquad E_\phi=\sqrt{m^2+g^2\bm\phi^2}.
\end{equation}
The eigenvalues are $-E_\phi,+E_\phi$, together with $n-1$ copies of $+m$ for the rest of charged flavors. Thus the fermions remain gapped for $m>0$. This also holds in a static nonuniform $\bm\phi$: writing $H=H_0+m\beta(P_v-P_0)$, the anticommutation of $P_v-P_0$ with every $T_I$ gives $H^2=H_0^2+m^2$.

The contact interaction needed in Section~\ref{on:HamON} also involves the density of the vector flavors. It is represented by a real scalar $\rho_v$:
\begin{equation}\label{on:vectordensitymother}
	S_{\rho_v}=\int\dd\tau\dd A\left[\frac{\rho_v^2}{4u_v}+\ii\rho_v\bar\Psi P_v\Psi\right],\qquad u_v>0.
\end{equation}
The field is omitted when $u_v=0$. Eliminating it and projecting to the LLL gives
\begin{equation}\label{on:vectordensitypositive}
	H_{\rho_v}=u_v\int\dd A:d_v^2:
	=2u_v\sum_{0<I<J}\int\dd A\,(\psi_J\psi_I)^\dagger(\psi_J\psi_I)\geq0,
\end{equation}
where $d_v=\sum_{I=1}^n\psi_I^\dagger\psi_I$. We also write $d_0=\psi_0^\dagger\psi_0$ and $n_{\rm tot}=d_0+d_v$. Normal ordering is understood throughout, with the one-body shifts absorbed into the flavor splitting. The special form of this contact channel will allow us to include it without changing the LLL coefficients computed below.

Other models use a total-density auxiliary instead:
\begin{equation}\label{on:densitymother}
	S_\rho=\frac12\rho V_\rho^{-1}\rho+\ii\int\dd\tau\dd A\,\rho\bar\Psi\Psi.
\end{equation}
It gives $H_\rho=\frac12:n_{\rm tot}V_\rho n_{\rm tot}:$. We specify which density channel is included in each model; the $O(n)$ model realization below uses $\rho_v$ only, the rotor models below use Eq.~\eqref{on:densitymother} instead. For a positive covariance $V_\rho$, $\rho$ is integrated over real values and has an imaginary coupling, as in Section~\ref{sub:twoorders}. Its precise effects on the neutral EFT depend on its kernel and will be stated later.

\subsection{The common neutral EFT in magnetic flux}\label{on:neutralEFT}
First set the density auxiliaries to zero and compute the $\bm\phi$ EFT for the mother \eqref{on:mother}. We then include the contact channel \eqref{on:vectordensitymother} in the strictly LLL theory. In the constant $\bm\phi$ background, the two active masses $\pm E_\phi$ give the same determinant as the Ising pair, with $\phi^2$ replaced by $\bm\phi^2$. Let $E_j=\sqrt{m^2+\lambda_j^2}$, where $\lambda_j$ is defined in \eqref{eq:diracLL}, and recall $d_j=Q+2j$. The change of potential is:
\begin{equation}\label{on:fluxpotential}
	\Delta V_Q=-\rho_0(E_\phi-m)-\frac{2}{\mathcal A}\sum_{j\geq1} d_j[\sqrt{E_j^2+g^2\bm\phi^2}-E_j]+\frac{\delta r}{2}\bm\phi^2.
\end{equation}
The counterterm removes the divergence in the quadratic coefficient. In the usual convention $V=r_Q\bm\phi^2/2+u_Q(\bm\phi^2)^2/4$, one obtains:
\begin{equation}\label{on:fluxcouplings}
	r_Q=r_0+\delta r-\frac{\rho_0g^2}{m}-\frac{2g^2}{\mathcal A}\sum_{j\geq1}\frac{d_j}{E_j},\qquad u_Q=\frac{\rho_0g^4}{2m^3}+\frac{g^4}{\mathcal A}\sum_{j\geq1}\frac{d_j}{E_j^3}>0.
\end{equation}

The above argument does not apply to terms with derivatives of $\bm\phi$. To obtain those, particularly the kinetic and stiffness corrections, we expand the full fermion determinant to quadratic order about $\bm\phi=0$. Each Yukawa vertex converts the singlet into flavor $I$, or vice versa. A loop with two vertices therefore requires the same flavor index $I$ at both vertices. Since all $n$ vector flavors have mass $+m$, each component $I$ gives exactly the Ising polarization, including its frequency and angular momentum dependence. This calculation allows arbitrary spacetime dependence of $\bm\phi$ and captures terms quadratic in $\bm\phi$, showing that they are diagonal in the flavor index $I$, with the kinetic and stiffness coefficients as in the Ising case. In particular, the LLL-generated polarization per flavor is \eqref{eq:LLLresponse}, including the LLL-generated kinetic and stiffness coefficients \eqref{eq:LLLkinetic} (see also Appendix \ref{app:spectral} and Supplement in \cite{MotherPRL}), which are relevant for the reverse route:
\begin{equation}\label{on:kinetic}
	Z_\tau^{\LLL}=\frac{\rho_0g^2}{4m^3},\qquad Z_s^{\LLL}=\frac{\rho_0g^2R^2}{mN}=\frac{g^2}{4\pi m}.
\end{equation}
For the Dirac mother without density interactions, the uniform HLL response \eqref{eq:HLLresponse} also gives
\begin{equation}\label{on:fullkinetic}
	Z_\tau=\frac{\rho_0g^2}{4m^3}+\frac{g^2}{2\mathcal A}\sum_{j\geq1}\frac{d_j}{E_j^3},\qquad Z_s=\kappa_s+\frac{g^2}{4\pi m}+\delta Z_s^{\HLL}.
\end{equation}
The spatial correction $\delta Z_s^\HLL$ includes both HLL--HLL and mixed LLL--HLL transitions and requires the nonuniform response; we do not evaluate it here. In the reverse route the charged fields are restricted to the LLL from the outset, so the HLL terms are absent.

The normalized occupied flavor eigenstate of eigenvalue $-E_\phi$ is proportional to:
\begin{equation}\label{on:realstate}
	|u(\bm\phi)\rangle\ \propto\ |0\rangle-\frac{g\phi_I}{m+E_\phi}|I\rangle.
\end{equation}
It can be chosen real and smooth for all $\bm\phi$. Consequently $\langle u|\partial_\tau u\rangle=0$: the occupied fermions generate no uniform Berry term. The occupied flavor subspace is one dimensional, with the real normalized basis vector displayed above. For $n=2$, the internal reflection also forbids $\epsilon_{II'}\phi_I\partial_\tau\phi_{I'}$, independently of the determinant calculation.

Now include $\rho_v$. At zero temperature and fixed filling $N_e=N$, integrating this channel in the normal-ordered LLL theory leaves the entire quadratic $\bm\phi$ response unchanged, for any $u_v\geq0$. At $\bm\phi=0$ the filled singlet has $N_v=\int\dd A\,d_v=0$. One Yukawa insertion creates a state with $N_v=1$. The interaction \eqref{on:vectordensitypositive} annihilates both $N_v=0$ and $N_v=1$ sectors, so it does not change the intermediate states in a two-point one-loop response, at any frequency or angular momentum. In particular, both coefficients in Eq.~\eqref{on:kinetic} remain valid after integrating over $\rho_v$.

The uniform LLL-induced potential is also unchanged. For a constant $\bm\phi=f\bm e$, all vector components of the filled state \eqref{on:realstate} occupy the same flavor direction $\bm e$. On this state the contact operator reduces to $:(\psi_{\bm e}^\dagger\psi_{\bm e})^2:=0$. As $H_{\rho_v}$ is positive and annihilates this ground state, it remains the ground state after adding the interaction $H_{\rho_v}$. Thus the full LLL potential is unchanged:
\begin{equation}\label{on:rhovpotential}
	\Delta V_{\LLL}=-\rho_0\left(\sqrt{m^2+g^2\bm\phi^2}-m\right).
\end{equation}
The contact interaction likewise annihilates the filled-shell flavor states reached by a spatially uniform time-dependent source $\bm\phi(t)=f(t)\bm{e}(t)$, and vanishes for a source of fixed flavor direction even if its length varies in spacetime, $\bm\phi=f(x,t)\bm{e}$, all for the same reason that $:(\psi_{\bm e}^\dagger\psi_{\bm e})^2:=0$. Its first effects require at least four $\phi$ insertions not of this form. At leading order in spatial derivatives, they can enter irrelevant interactions such as:
\begin{equation}\label{on:rhovderivatives}
	\bm\phi^2(\boldsymbol\nabla\bm\phi)^2-\sum_i(\bm\phi\cdot\nabla_i\bm\phi)^2 =f^4(\boldsymbol\nabla\bm e)^2,
\end{equation}
whose coefficient is not needed for the leading EFT. This conclusion is only about the LLL effects, and thus the reverse route. It does not include the HLL-induced effects relevant to the full UV complete Dirac mother, whose HLL sea and spinor components must be included separately.

The leading neutral EFT therefore has the form
\begin{equation}\label{on:eft}
	\mathcal L_{O(n)}=\frac{Z_\tau}{2}(\partial_\tau\bm\phi)^2+\frac{Z_s}{2}(\boldsymbol{\nabla}\bm\phi)^2+\frac{r_Q}{2}\bm\phi^2+\frac{u_Q}{4}(\bm\phi^2)^2+\cdots.
\end{equation}
When $Z_\tau,Z_s>0$, the time rescaling $v^2=Z_s/Z_\tau$ brings the kinetic term to relativistic form. With a positive quartic and a tuned mass this is the usual $O(n)$ Wilson--Fisher EFT~\cite{WilsonFisher}. The coefficients above precede fluctuations of $\bm\phi$ itself, which still determine the physical critical coupling. Such couplings are straightforward to find numerically from the Hamiltonian considered next, whose derivative interaction around the contact ground states provides an explicit regime in which the Wilson--Fisher flow is accessible.

\subsection{The general $O(n)$ fuzzy Hamiltonian}\label{on:HamON}

The construction of Ref.~\cite{GuoON} uses the same $n+1$ flavors at $N_e=N$, with the interaction:
\begin{equation}\label{on:guo}
	H_{\rm WF}=\int\dd A\left[u_0:n_{\rm tot}^2:-U_1\sum_I:V_I\boldsymbol\nabla^2V_I:+2h_1d_0\right],\qquad V_I=\psi^\dagger T_I\psi,
\end{equation}
with $u_0,U_1>0$. While in \cite{GuoON} $u_0=1$, we display it to separate the charge and anisotropy scales. All couplings here include the conversion to our area-normalized fields. The coefficient $h_1$ also includes the one-body correction from normal ordering. Indeed, $\sum_IT_I^2=nP_0+P_v$, so, after the normalization conversion, comparison with \cite{GuoON} is:
\begin{equation}\label{on:guoordering}
	h_1=h_1^{\rm their}+\frac{U_1\rho_0(N-1)(n-1)}{2R^2}.
\end{equation}
The other contractions only shift the energy zero. The density term is part of this Hamiltonian and is responsible for its contact ferromagnetic behavior when $U_1=h_1=0$.

The same $\phi_I$ mother \eqref{on:mother} reproduces this Hamiltonian once its contact interaction is completed by the $\rho_v$ channel \eqref{on:vectordensitymother}. The needed LLL identities are:
\begin{equation}\label{on:contactidentity}
	:\!\sum_I V_I^2\!:\,=-2:d_0d_v:,\qquad
	:n_{\rm tot}^2:=2:d_0d_v:+:d_v^2:.
\end{equation}
Here $:d_0^2:=0$ by Pauli exclusion, whereas $:d_v^2:=2\sum_{I<J}:d_Id_J:$ need not vanish. Thus the $\phi_I$ contact interaction already supplies the singlet--vector repulsion, and $\rho_v$ supplies the remaining repulsion between distinct vector flavors. For $n=1$ the latter vanishes. This is a consequence of the contact flavor algebra; no choice of the spatial kernel for the $T_I$ channels can generate the missing interaction between two distinct vector flavors, which is why the $\rho_v$ channel is necessary.

Explicitly, take the mother $S_{O(n)}+S_{\rho_v}$ of Eqs.~\eqref{on:mother} and \eqref{on:vectordensitymother}, with
\begin{equation}\label{on:guomatching}
	g^2K_0=2(u_0+U_1\boldsymbol\nabla^2),\qquad u_v=u_0,\qquad m=-h_1.
\end{equation}
Eliminating the mediators and projecting gives:
\begin{equation}\label{on:guoelimination}
	\begin{aligned}
		H_\phi&=\int\dd A\left[2u_0:d_0d_v:-U_1\sum_I:V_I\boldsymbol\nabla^2V_I:\right],\\
		H_{\rho_v}&=u_0\int\dd A:d_v^2:.
	\end{aligned}
\end{equation}
Their sum is the interaction in Eq.~\eqref{on:guo}. The splitting obeys $m(N_v-N_0)=mN_e-2mN_0$, so the last relation in Eq.~\eqref{on:guomatching} holds up to a fixed-number constant and uses the normal-ordering convention already stated. The singlet side has $h_1<0$, as required by $m>0$. The harmonic eigenvalues are $K_{0,L}=2[u_0-U_1L(L+1)/R^2]/g^2$. Any negative eigenvalues are represented by the Gaussian source phases; modes of zero covariance are omitted. The matching is exact in the LLL reverse route. A Dirac completion requires the additional HLL matching described above.

The density term in Ref.~\cite{GuoON} was included to suppress local charge fluctuations and reduce finite-size effects. In our representation part of it comes from the contact piece of $K_0$, and the rest from $\rho_v$. The choice of a pure Laplacian for the ultralocal interaction in \cite{GuoON} is therefore consistent with our contact plus Laplacian HS representation.

\subsubsection{Yet another path to the EFT}
For weak $U_1$, we can derive the neutral EFT by expanding about the exact polarized ground states of the contact interaction. This includes the collective dynamics left to be integrated in Section~\ref{on:neutralEFT}.

At $U_1=h_1=0$, the contact Hamiltonian has the ground states
\begin{equation}\label{on:cpground}
	|F_z\rangle=\prod_{m=-s}^s\left(\sum_{a=0}^nz_ac_{ma}^\dagger\right)|0\rangle,\qquad z^\dagger z=1,\qquad z\sim e^{\ii\gamma}z.
\end{equation}
Pauli exclusion makes their contact energy vanish. Their flavor orientations form $\mathbb{CP}^n$, parametrized locally by a slowly varying collective field $z(\mathbf r,\tau)$. We assume a nonzero charge gap $\Delta_c$ and retain the soft neutral modes at a finite Wilsonian cutoff. Short range interactions and separation from the eliminated modes then permit an expansion in derivatives. Define:
\begin{equation}\label{on:tangentderivative}
	D_\mu z=(1-zz^\dagger)\partial_\mu z,\qquad z^\dagger D_\mu z=0.
\end{equation}
This projects the spacetime derivative onto the tangent space of $\mathbb{CP}^n$, removing unphysical directions. The contact theory's $SU(n+1)$ symmetry fixes the quadratic form on this tangent space up to a constant. Spatial rotations contract its spatial indices, while the temporal coefficient is independent. Retaining these derivative terms and the leading anisotropy potential gives:
\begin{equation}\label{on:zaction}
	\mathcal L_z=\rho_0z^\dagger\partial_\tau z+\frac{\mathcal J}{2}|D_i z|^2+\frac{\chi}{2}|D_\tau z|^2+\mathcal E(z)+\cdots.
\end{equation}
The ellipsis includes higher derivatives and anisotropy corrections to the derivative terms. A spatial topological term is constant in the sector of smooth fluctuations about uniform $z$ and is omitted. The overlap $\langle F_{z'}|F_z\rangle=(z'^\dagger z)^N$ fixes the Berry coefficient to $\rho_0=N/\mathcal A$. The term with coefficient $\chi$ is allowed; it is subleading at the contact ferromagnet, where $\omega\sim k^2$, but we keep it to track its effect. The exact contact one-magnon dispersion fixes the spatial coefficient:
\begin{equation}\label{on:contactdispersion}
	\epsilon_L=2u_0\rho_0[1-p_L(N)]\simeq\frac{u_0}{2\pi}k_L^2,\qquad k_L^2=\frac{L(L+1)}{R^2},\qquad \mathcal J=\frac{u_0\rho_0}{\pi}.
\end{equation}
For $L(L+1)\ll N$, $p_L(N)=1-L(L+1)/N+\cdots$. Equation~\eqref{on:zaction} gives $\epsilon(k)=\mathcal Jk^2/(2\rho_0)+O(k^4)$, fixing the displayed large-$N$ value of $\mathcal J$.

The potential is the uniform-state energy density $\mathcal E(z)=\langle F_z|H_{\rm WF}|F_z\rangle/\mathcal A$ for a fixed $z$. Wick contraction gives, for any Hermitian flavor matrix $T$,
\begin{equation}\label{on:wick}
	\langle:(\psi^\dagger T\psi)(x)(\psi^\dagger T\psi)(y):\rangle=(z^\dagger Tz)^2[\rho_0^2-|\mathcal P(x,y)|^2],\qquad |\mathcal P(x,y)|^2=\rho_0^2\left(\frac{1+\widehat x\cdot\widehat y}{2}\right)^{N-1}.
\end{equation}
The two-point function depends on the separation between $x$ and $y$ through the exchange term $|\mathcal P(x,y)|^2$. The Laplacian $\nabla_y^2$ acts on this term before setting $y=x$, giving a nonzero derivative interaction contribution to the energy proportional to $\rho_0^2(N-1)/R^2$:
\begin{equation}\label{on:guopotential}
	\mathcal E(z)=-A_1\sum_I(z^\dagger T_Iz)^2+2h_1\rho_0|z_0|^2,\qquad A_1=\frac{U_1\rho_0^2(N-1)}{R^2}>0.
\end{equation}
This is the leading potential for weak anisotropy; collective fluctuations remain to be integrated. We neglect the relative $O(U_1B/u_0)$ corrections to the spatial stiffness.

To expand near the singlet orientation, choose $z_0$ real and write:
\begin{equation}\label{on:xycoordinates}
	z_0=\sqrt{1-\bm x^2-\bm y^2},\qquad z_I=x_I+\ii y_I,\qquad z^\dagger T_Iz=2x_I\sqrt{1-\bm x^2-\bm y^2}.
\end{equation}
Thus $\langle V_I\rangle=2\rho_0x_I+\cdots$ identifies $\bm x$ as the real $O(n)$ order parameter, related to the mediator by $\langle\phi_I\rangle=-gK_0\langle V_I\rangle$. The Berry term becomes:
\begin{equation}\label{on:xyberry}
	z^\dagger\partial_\tau z=\ii(\bm x\cdot\partial_\tau\bm y-\bm y\cdot\partial_\tau\bm x)=-2\ii\bm y\cdot\partial_\tau\bm x+\ii\partial_\tau(\bm x\cdot\bm y).
\end{equation}
Dropping the total derivative and expanding the derivative terms to quadratic order gives:
\begin{equation}\label{on:guolocalEFT}
	\begin{aligned}
		\mathcal L={}&-2\ii\rho_0\bm y\cdot\partial_\tau\bm x+\frac{\chi}{2}[(\partial_\tau\bm x)^2+(\partial_\tau\bm y)^2]+\frac{\mathcal J}{2}[(\nabla\bm x)^2+(\nabla\bm y)^2]\\
		&-(4A_1+2\rho_0h_1)\bm x^2-2\rho_0h_1\bm y^2+4A_1\bm x^2(\bm x^2+\bm y^2)+\cdots.
	\end{aligned}
\end{equation}
At $h_{1,*}=-2A_1/\rho_0$, the $\bm x^2$ coefficient vanishes while the $\bm y^2$ coefficient remains $4A_1>0$. For $\mathcal Jk^2,|\chi|\omega^2\ll8A_1$, (so $k^2\ll k_*^2=8A_1/\mathcal{J}$,) the derivatives of $\bm y$ are small corrections, and its Gaussian integral follows by completing the square:
\begin{equation}\label{on:ysquare}
	4A_1\bm y^2-2\ii\rho_0\bm y\cdot\partial_\tau\bm x=4A_1\left(\bm y-\frac{\ii\rho_0}{4A_1}\partial_\tau\bm x\right)^2+\frac{\rho_0^2}{4A_1}(\partial_\tau\bm x)^2.
\end{equation}
The mixed quartic produces nonlinear derivative terms at tree level. Allowing a small mass detuning, we obtain:
\begin{equation}\label{on:guocouplings}
	\begin{aligned}
		\mathcal L_{\bm x}&=\frac{K_t}{2}(\partial_\tau\bm x)^2+\frac{K_s}{2}(\boldsymbol\nabla\bm x)^2+\frac r2\bm x^2+4A_1(\bm x^2)^2+\cdots,\\
		K_t&=\chi+\frac{\rho_0^2}{2A_1},\qquad K_s\simeq\mathcal J,\qquad r=-8A_1-4\rho_0h_1.
	\end{aligned}
\end{equation}
We do not compute $\chi$. If the eliminated modes remain separated from zero energy as $U_1\to0$ at a fixed cutoff, it is regular, and sufficiently weak anisotropy gives $2A_1|\chi|/\rho_0^2\ll1$. The Berry contribution then dominates $K_t$. We get the usual $O(n)$ scalar action with positive kinetic and quartic coefficients. Collective fluctuations renormalize these coefficients and shift the critical field.

Rescaling time with $v^2=K_s/K_t$ and setting $\bm\varphi=(K_tK_s)^{1/4}\bm x$, the quartic $\lambda_4(\bm\varphi^2)^2/4!$ and spatial crossover scale $k_*$ obey
\begin{equation}\label{on:guowindow}
	\lambda_4=\frac{96A_1}{\sqrt{K_t}K_s^{3/2}},\qquad k_*^2=\frac{8A_1}{\mathcal J},\qquad \frac{\lambda_4}{k_*}\simeq\frac{48A_1}{\rho_0\mathcal J}.
\end{equation}
The last estimate uses $2A_1|\chi|/\rho_0^2\ll1$. At fixed $B,n$, sufficiently small $U_1B/u_0$ and large $R$ permit $R^{-1}\ll\lambda_4\ll k_*$, with $vk_*\simeq4A_1/\rho_0\ll\Delta_c,\Delta_{\rm LL}$. Retuning the mass then gives the $O(n)$ Wilson--Fisher flow.

\subsection{The integer $O(2)$ rotor}

The three charged flavors of Ref.~\cite{DeyOtwo} carry internal spin one, $q=+1,0,-1$. This spin is unrelated to the spatial angular momentum of the LLL. In the Cartesian basis
\begin{equation}\label{on:spinonebasis}
	|x\rangle=\frac{|+\rangle+|-\rangle}{\sqrt2},\qquad |y\rangle=\frac{\ii(|+\rangle-|-\rangle)}{\sqrt2},
\end{equation}
one has $S^x=T_x$, $S^y=-T_y$, and $(S^z)^2=P_v$. Thus the mother is explicitly Eq.~\eqref{on:mother} at $n=2$, with the mass matrix
\begin{equation}\label{on:spinonemother}
	\mathcal M=m[2(S^z)^2-\mathbf1]+g(\phi_xS^x+\phi_yS^y).
\end{equation}
Its internal $O(2)$ consists of rotations generated by $S^z$ and a reflection exchanging the $q=+1$ and $q=-1$ sectors. This reflection leaves the occupied $q=0$ LLL invariant and forbids the quadratic Berry term $\ii (\phi_x\partial_\tau \phi_y - \phi_y\partial_\tau\phi_x)$.

Write $\mathcal U$ for the spatial kernel whose LLL matrix elements are used in that paper. Its Hamiltonian and mother parameters are
\begin{equation}\label{on:o2published}
	H_{O(2)}=:n_{\rm tot}\mathcal U n_{\rm tot}:-\frac12\sum_{I=x,y}:J_I\mathcal UJ_I:+D_{\rm an}N_v,\qquad J_I=\psi^\dagger S^I\psi,
\end{equation}
\begin{equation}\label{on:o2matching}
	g^2K_0=\mathcal U,\qquad V_\rho=2\mathcal U,\qquad 2m=D_{\rm an}.
\end{equation}
The last equality includes any shift required on converting the operator products in \cite{DeyOtwo} to normal ordering. These equations specify the density auxiliary \eqref{on:densitymother} as well as the two scalar mediators. If $\mathcal U_L$ changes sign, each harmonic is represented by a real Gaussian of covariance $|\mathcal U_L|$, with a real or imaginary source coupling as required by the interaction sign; there is no change of integration contour. The common calculation of \eqref{on:eft} applies in this case, the actual density and exchange couplings determining its coefficients. The phase transition of Hamiltonian \eqref{on:o2published}, and its $O(2)$ spectrum, are established numerically in \cite{DeyOtwo}.

\subsection{The $O(3)$ singlet--triplet rotor}

A quantum rotor is a particle whose coordinate is a unit vector on an internal sphere. Its wavefunctions have angular momentum $\ell=0,1,\ldots$. Keeping $\ell=0,1$ on the internal $S^2$ leaves one singlet and three triplet states. These four states become the four charged flavors of Ref.~\cite{DeyOthree} (note: this internal sphere is different from the spatial $S^2$ that is made fuzzy). If $P_{01}$ projects onto this internal truncation, then:
\begin{equation}\label{on:rotormat}
	P_{01}\widehat n_IP_{01}=\frac{T_I}{\sqrt3},\qquad P_{01}\widehat{\bm L}^{\,2}P_{01}=2P_v,\qquad I=1,2,3.
\end{equation}
Consequently its mother is Eq.~\eqref{on:mother} at $n=3$, together with the density auxiliary. More explicitly, with $n_I=\psi^\dagger T_I\psi/\sqrt3$,
\begin{equation}\label{on:o3matching}
	H_{O(3)}=u:n_{\rm tot}\mathcal U n_{\rm tot}:-v\sum_I:n_I\mathcal U n_I:+2hN_v,\qquad g^2K_0=\frac{2v}{3}\mathcal U,\quad V_\rho=2u\mathcal U,\quad m=h.
\end{equation}
Here $h$ denotes the singlet--triplet splitting after the normal-ordering correction has been included. This fixes its complete Gaussian auxiliary action through Eqs.~\eqref{on:mother} and \eqref{on:densitymother}. The quarter filling is $N_e=N$. Internal $O(3)$ rotates the triplet $\Psi_I$ and the neutral scalar triplet $\phi_I$ together; inversion changes their signs and leaves the singlet $\Psi_0$ invariant. The $\phi_I$ EFT is the three-component instance of Eq.~\eqref{on:eft}, with zero Berry term and a positive induced quartic. The critical couplings are determined by the numerical analysis in \cite{DeyOthree}. The rotor model and \eqref{on:guo} at $n=3$ have the same degrees of freedom and symmetry but different spatial interactions.

\subsection{The Heisenberg bilayer}

The bilayer of \cite{HanBilayer} uses four charged flavors $\Psi_{\ell\sigma}$, with layer $\ell=1,2$, internal spin $\sigma=\uparrow,\downarrow$. The Pauli matrices acting on these indices are denoted $\tau^a$ and $\sigma^I$, respectively. In the mother theory, as usual, we promote $\Psi_{\ell\sigma}$ to Dirac spinors (with the spinor index suppressed), and write:
\begin{equation}\label{on:bilayermother}
	S_{\rm BL}=\int\dd\tau\dd A\,\bar\Psi[\slashed D_A-m\tau^x+g\phi_I\tau^z\sigma^I+g_b b_I\sigma^I+\ii\rho]\Psi+\frac12\phi_IK_-^{-1}\phi_I+\frac12b_IK_+^{-1}b_I+\frac12\rho V_\rho^{-1}\rho.
\end{equation}
Here $\bm\phi$ couples to the staggered spin (the difference of two layers), while $\bm b$ -- to the uniform spin (the sum of two layers). Both are triplets under the internal spin rotations. Layer exchange acts by $\tau^x$, sends $\bm\phi\mapsto-\bm\phi$, and leaves $\bm b$ unchanged. The total charged $U(1)$ and spatial rotations are also preserved. The low energy $O(3)$ EFT of $\bm\phi$ requires the $\bm b$ and $\rho$ fields to remain heavy.

At $\bm b=\rho=0$, define the $4\times 4$ flavor mass matrix:
\begin{equation}
	\mathcal{M}(\phi)=-m\tau^x \otimes \mathbf{1} + g \tau^z\otimes (\bm\phi \cdot \bm\sigma).
\end{equation}
The four matrices $\tau^x\otimes \mathbf{1},\tau^z\otimes\sigma^I$ anticommute and form a Clifford algebra. The matrix satisfies $\mathcal{M}^2=E_\phi^2$, where $E_\phi=\sqrt{m^2+g^2\bm\phi^2}$. The mass eigenvalues are $\pm E_\phi$, each twice degenerate. Both negative branches are filled: $N_e=2N$, in contrast to the four-flavor rotor at $N_e=N$. The potential and kinetic coefficients from the charged determinant are twice those of Eqs.~\eqref{on:fluxcouplings} and \eqref{on:kinetic}. The occupied projector gives:
\begin{equation}\label{on:bilayerberry}
	P_-=\frac12(\mathbf1-\mathcal M/E_\phi),\qquad \mathcal F_{II'}=\ii\Tr\{P_-[\partial_I P_-,\partial_{I'}P_-]\}=0.
\end{equation}
The trace vanishes because a product of three distinct matrices in this Clifford representation has zero trace. Thus the two occupied branches generate no quadratic Berry term for $\bm\phi$. Eliminating the massive uniform-spin response modifies the coefficient of $(\partial_\tau\bm\phi)^2$; it does not generate the Berry term either. The remaining action has the form \eqref{on:eft} for three real components.

Define $\bm n_{1,2}=\psi^\dagger[(\mathbf1\pm\tau^z)/2]\bm\sigma\psi$ and $\bm N_\pm=\bm n_1\pm\bm n_2$. Gaussian elimination of $\bm\phi, \bm{b}, \rho$ gives:
\begin{equation}\label{on:bilayerreverse}
	H_{\rm BL}=\frac12:n_{\rm tot}V_\rho n_{\rm tot}:-\frac12g^2:\bm N_-K_-\bm N_-:-\frac12g_b^2:\bm N_+K_+\bm N_+:-m\int\dd A\,\psi^\dagger\tau^x\psi.
\end{equation}
The choices
\begin{equation}\label{on:bilayerkernels}
	V_\rho=2U_0,\qquad g^2K_-=U_1+\frac12U_2,\qquad g_b^2K_+=U_1-\frac12U_2
\end{equation}
give $U_0n_{\rm tot}n_{\rm tot}-U_1(\bm n_1\bm n_1+\bm n_2\bm n_2)+U_2\bm n_1\bm n_2$, with the spatial kernels implicit, precisely the interaction used in \cite{HanBilayer}. Their choice $U_2\propto\boldsymbol\nabla^2\delta$ need not give positive harmonic eigenvalues for both $K_\pm$; the Gaussian source phases implement those signs. The density and uniform-spin corrections at the numerical critical point follow from the results in \cite{HanBilayer}, whose numerics supply the evidence for the infrared $O(3)$ identification. We should also emphasize that the bilayer model (with its mother) is different from the earlier $O(3)$ realizations we had, even though they describe the same CFT.

\subsection{The cubic model and its mother}

The cubic model of Ref.~\cite{StergiouCubic} starts from the four-flavor singlet--triplet rotor, with $n=3$, and adds a two-body interaction among the three Cartesian triplet densities. Define $P_I=|I\rangle\langle I|$, $d_I=\psi^\dagger P_I\psi$, and use the convention
\begin{equation}\label{on:cubham}
	H_{\rm cub}=\frac w2\sum_{I=1}^3:d_I\mathcal{U} d_I:.
\end{equation}
The cubic deformation uses the same spatial kernel $\mathcal U$ as the $O(3)$ rotor interaction, with an independent coupling $w$. The mother is explicitly the rotor action plus three scalar density auxiliaries:
\begin{equation}\label{on:cubHS}
	S_{{\rm cubic},M}=S_{O(3)}+S_\rho+\frac12\eta_I\mathcal{U}^{-1}\eta_I+\ii\sqrt w\int\dd\tau\dd A\,\eta_I\bar\Psi P_I\Psi.
\end{equation}
For $w>0$ and positive $\mathcal{U}$, the $\eta_I$ have real integration domains. Signed kernel eigenvalues are handled as above. The remaining symmetry is the cubic group of permutations and sign flips: permutations act simultaneously on $\Psi_I$, $\phi_I$, and $\eta_I$; independently, each pair $(\Psi_I,\phi_I)$ may change sign, with $\eta_I$ unchanged. Eliminating $\eta_I$ gives \eqref{on:cubham}, and eliminating $\bm\phi,\rho$ gives the other rotor interactions. The cubic normal-ordering correction is proportional to $P_v$ and is included in the singlet--triplet splitting.

To relate our projector interaction to the schematic rotor deformation $\sum_I\hat n_I^{\,2}\otimes\hat n_I^{\,2}$, project each squared orientation operator onto the singlet and triplet states:
\begin{equation}\label{on:cubprojectors}
	Q_I=P_{01}\widehat n_I^{\,2}P_{01}=\frac13P_0+\frac15P_v+\frac25P_I.
\end{equation}
Inserting Eq.~\eqref{on:cubprojectors} into $\sum_I Q_I\otimes Q_I$ gives $(4/25)\sum_IP_I\otimes P_I$ plus $O(3)$-invariant density terms, and \cite{StergiouCubic} already states that and adopts the projector convention, retuning the invariant couplings accordingly.

We can compute the induced cubic anisotropy perturbatively in $w$. In the uniform occupied eigenstate \eqref{on:realstate}, the probability of flavor $I$ is:
\begin{equation}\label{on:pa}
	\langle u(\phi)|P_I|u(\phi)\rangle=\frac{g^2\phi_I^2}{2E_\phi(E_\phi+m)}=\frac{g^2\phi_I^2}{4m^2}+O(\phi^4).
\end{equation}
The leading perturbative potential is $\Delta V=\frac{\langle H_{\rm cub}\rangle}{\mathcal A}$, where the average is taken in a state obtained by filling every LLL orbital with \eqref{on:realstate}. Applying the same contractions as in Eq.~\eqref{on:wick} gives:
\begin{equation}\label{on:IW}
	\mathcal I_{\mathcal{U}}=\frac1{\mathcal A}\int\dd A_x\dd A_y\,\mathcal{U}(x,y)[\rho_0^2-|\mathcal P(x,y)|^2],\qquad \Delta V=\frac{wg^4\mathcal I_{\mathcal{U}}}{32m^4}\sum_I\phi_I^4+O(w\phi^6,w^2).
\end{equation}
This coefficient vanishes for a strictly contact interaction between identical flavors. More generally, $\mathcal I_{\mathcal{U}}$ is twice the trace of the projected two-particle kernel on the antisymmetric orbital Hilbert space, divided by $\mathcal A$. Nonnegative odd pseudopotentials $V_{2k+1}$ therefore give $\mathcal I_{\mathcal{U}}\geq0$, with strict positivity when $V_1>0$.

The neutral EFT is therefore
\begin{equation}\label{on:cubicEFT}
	\mathcal L_{\rm cub}=\frac{Z_\tau}{2}(\partial_\tau\bm\phi)^2+\frac{Z_s}{2}(\nabla\bm\phi)^2+\frac r2\bm\phi^2+\frac u4(\bm\phi^2)^2+\frac v4\sum_I\phi_I^4+\cdots,\qquad v=\frac{wg^4\mathcal I_{\mathcal{U}}}{8m^4}+O(w^2).
\end{equation}
Independent component reflections forbid a quadratic Berry term. The allowed terms with one time derivative at quartic order are total derivatives. Thus the weak-deformation calculation gives the usual relativistic cubic action, with higher derivative corrections. The quartic is bounded below when $u+v>0$ and $u+v/3>0$. A mass tuning in the stable cubic basin gives the cubic fixed point \cite{Aharony1973Cubic,CarmonaPelissettoVicari2000}. The expression for $v$ is the leading term in the expansion around $w=0$, it does not determine the fully renormalized coupling. Reference \cite{StergiouCubic} supplies the numerical identification of couplings in the Hamiltonian.

\subsection{Free scalars and the second $O(n)$ Hamiltonian}\label{on:freevector}

The second Hamiltonian of Ref.~\cite{GuoON} uses the same $n+1$ charged flavors:
\begin{equation}\label{on:freepublished}
	H_2=\int\dd A\left[u_0:n_{\rm tot}^2:+U_2\sum_{0\leq\alpha<\beta\leq n}:\omega_{\alpha\beta}\nabla^2\omega_{\alpha\beta}:+2\mu d_0\right].
\end{equation}
Here $\omega_{\alpha\beta}=\ii(\psi_\alpha^\dagger\psi_\beta-\psi_\beta^\dagger\psi_\alpha)$, $d_0=\psi_0^\dagger\psi_0$, and $\mu=h_2/R^2$ for the published pinning prescription. Rephasing $\psi_0=\ii\widetilde\psi_0$ turns $\omega_{0I}$ into $V_I$. Thus, with $H_1$ denoting Eq.~\eqref{on:guo} and its splitting set to $h_1=\mu$,
\begin{equation}\label{on:freecomparison}
	H_2(U_2,\mu)=H_1(-U_2,\mu)+U_2\sum_{1\leq I<J\leq n}\int\dd A:\omega_{IJ}\nabla^2\omega_{IJ}:.
\end{equation}
For $n=1$ the extra sum is absent. Each extra channel admits the Gaussian auxiliary representation of Section~\ref{sec:framework}.

We use the previously derived EFT for the flavor orientation $z$ of the contact ground state \eqref{on:cpground}, treating the $U_2$ interaction as a small perturbation and assuming that the charge gap remains nonzero. The uniform occupied vector $z^\dagger z=1$ gives, to first order in $U_2$, energy density:
\begin{equation}\label{on:freeanisotropy}
	\mathcal E(z)=A_2[1-|z^Tz|^2]+2\rho_0\mu|z_0|^2,\qquad A_2=\frac{U_2\rho_0^2(N-1)}{R^2}.
\end{equation}
This follows from evaluating the interaction in $|F_z\rangle$ using Wick’s theorem, as in Eq.~\eqref{on:wick}, and using the identity $\sum_{\alpha<\beta}(z^\dagger T_{\alpha\beta}z)^2=1-|z^Tz|^2$, where $T_{\alpha\beta}=\ii(|\alpha\rangle\langle\beta|-|\beta\rangle\langle\alpha|)$.

For $U_2>0$ and $\mu=0$, the minima are real vectors up to phase, with $z\sim-z$: the ordered manifold is $\mathbb{RP}^n$, giving $n$ Goldstones. Write $z=(\sqrt{1-\bm x^2-\bm y^2},\bm x+\ii\bm y)$. The Berry term $-2\ii\rho_0\bm y\cdot\partial_\tau\bm x$ and the energy $4A_2\bm y^2$ give, after eliminating $\bm y$,
\begin{equation}\label{on:freeinertia}
	\mathcal L_x=\frac{K_t}{2}(\partial_\tau\bm x)^2+\frac{K_s}{2}(\nabla\bm x)^2-\frac{2\rho_0h_2}{R^2}\bm x^2+\cdots,\qquad K_t=\frac{\rho_0^2}{2A_2},\quad K_s\simeq\mathcal J.
\end{equation}
As above, other kinetic corrections are subleading at small $U_2$. Choosing $h_2=-K_s/(16\rho_0)$ gives the conformal mass and $E_L=\sqrt{K_s/K_t}(L+1/2)/R$ at this order. Nonlinear corrections vanish when $R\sqrt{K_tK_s}\gg1$, provided charged excitations and projective defects remain above these levels. This is the ordered-phase Goldstone construction of free scalars, similar to Section \ref{sec:goldstone}.

Negative $U_2$ instead gives an interacting critical regime. Now $A_2<0$, and $\mathcal{E}(z)$ at parameter $\mu_*=2A_2/\rho_0$ gives $\bm x$ positive energy $-4A_2\bm x^2$, while $\bm y$ becomes soft. Eliminating $\bm x$ gives:
\begin{equation}\label{on:freeWF}
	\mathcal L_y=\frac{\rho_0^2}{4|A_2|}(\partial_\tau\bm y)^2+\frac{\mathcal J}{2}(\nabla\bm y)^2+\frac r2\bm y^2+4|A_2|(\bm y^2)^2+\cdots,\qquad r=-8|A_2|-4\rho_0\mu.
\end{equation}
The positive quartic gives the $O(n)$ Wilson--Fisher flow for sufficiently small $|A_2|/(\rho_0\mathcal J)$, after mass retuning. This predicts a critical regime beyond the published positive $U_2$ parameters; $\mu=\mu_*$ is its leading estimate, and reaching it requires $h_2\sim R^2$. Near this transition, the extra channels in Eq.~\eqref{on:freecomparison} are expected to be irrelevant in the RG sense. Their irrelevance there does not justify dropping them at positive $U_2$, where they help enforce the real projective vacuum manifold.

\section{The three-state Potts model}\label{pc:potts}

The model from \cite{pc:YangPotts} uses three fermion flavors at one-third filling of the LLL. It has an internal $S_3$ permutation symmetry and a transition between a symmetric phase and a phase that selects one color. The reported spectra are approximately conformal, but their finite size drift favors a weak first order transition. Our task here is to recover the scalar EFT and the published Hamiltonian. Neither calculation proves or disproves a real Potts fixed point in three dimensions.

\subsection{Three charged flavors and an $S_3$ order parameter}

Let $\Psi=(\Psi_0,\Psi_1,\Psi_2)^T$ consist of three charge-one Dirac fermions. The following matrices act on this flavor space:
\begin{equation}\label{pc:clock}
	\omega=e^{2\pi\ii/3},\qquad Z=\operatorname{diag}(1,\omega,\omega^2),\qquad X=\begin{pmatrix}0&0&1\\1&0&0\\0&1&0\end{pmatrix}.
\end{equation}
A complex scalar $\Phi$ couples to the color density $J_Z=\bar\Psi Z\Psi$. A real auxiliary $\rho$ supplies the total-density interaction, with $J_0=\bar\Psi\Psi$. The mother action is:
\begin{align}\label{pc:tripletmother}
	S_3={}&\int\dd\tau\,\dd A\,\bar\Psi\left[\slashed D_A+\mathsf M(\Phi)+\ii\rho\right]\Psi +\Phi^*K^{-1}\Phi+\frac12\rho K_\rho^{-1}\rho,\\
	\mathsf M(\Phi)={}&m_0\mathbf1-h(X+X^\dagger) +g(\Phi^*Z+\Phi Z^\dagger),\qquad h>0. \label{pc:tripletmass}
\end{align}
Kernel products include the spacetime integrals. We take them instantaneous as before, and their spatial eigenvalues will be fixed below. The integration over $\rho$ is along the real contour and its imaginary linear coupling gives a repulsive density term. Besides spatial rotations and electric charge, the theory is invariant under a cyclic transformation $\mathsf C_3$ and a flavor reflection $\mathsf R$:
\begin{align}\label{pc:permutations}
	\mathsf{C}_3:\quad &\Psi\mapsto X\Psi,\qquad\Phi\mapsto\omega\Phi;\cr \mathsf{R}:\quad &\Psi\mapsto R\Psi,\qquad\Phi\mapsto\Phi^*, \qquad R=\begin{pmatrix}1&0&0\\0&0&1\\0&1&0\end{pmatrix},
\end{align}
which generate $S_3$ and leave $\rho$ invariant.

At $\Phi=0$, the symmetric flavor state $|s\rangle=(|0\rangle+|1\rangle+|2\rangle)/\sqrt3$ has mass $m_0-2h$, while its two orthogonal partners have mass $m_0+h$. At this Gaussian reference point, choosing
\begin{equation}\label{pc:masswindow}
	-h<m_0<2h,\qquad N_e=N,\qquad \Delta_c=\min(2h-m_0,m_0+h)>0
\end{equation}
fills the singlet LLL and leaves the other two flavors empty. The one-body term $m_0N_e$ is constant in this fixed-filling Hamiltonian; its role in the mother theory is to select the charged vacuum. The gap between singlet and doublet states is:
\begin{equation}\label{pc:tripletgap}
	\Delta=3h.
\end{equation}
A small uniform $\Phi$ leaves the charged gap open: the perturbation of $\mathsf M$ has operator norm at most $2|g\Phi|$, so $2|g\Phi|<\Delta_c$ is a sufficient condition, while $2|g\Phi|\ll \Delta_c$ is required for the local EFT. The neutral EFT also requires the density mediator $\rho$ to remain noncritical. We compute the charged determinant at a uniform density saddle $\rho_*$, absorbing its constant contribution into $m_0$. Integrating its fluctuations would give additional renormalizations, not included in the coefficients displayed below.

\subsection{The scalar EFT in magnetic flux}

The uniform LLL contribution can be found by diagonalizing the $3\times 3$ flavor matrix. Write its lowest eigenvalue as $m_0-2h+E(\Phi)$. Its characteristic equation is:
\begin{equation}\label{pc:cubiccharacteristic}
	E(E-\Delta)^2-g^2|\Phi|^2(3E-2\Delta)-g^3(\Phi^3+\Phi^{*3})=0.
\end{equation}
Expanding the root with $E(0)=0$ gives the LLL contribution to the effective potential per unit area:
\begin{equation}\label{pc:potential}
	\delta v_{\LLL}=\rho_0\left[-\frac{2g^2}{\Delta}|\Phi|^2 +\frac{g^3}{\Delta^2}(\Phi^3+\Phi^{*3}) +\frac{2g^4}{\Delta^3}|\Phi|^4+O(\Phi^5)\right].
\end{equation}
Thus the cubic Potts invariant is present in the charged determinant. It is allowed by $S_3$ and cannot generally be omitted.

The two virtual transitions from the occupied singlet contribute with opposite frequency signs. Using the LLL form factor $p_L(N)$ defined in \eqref{eq:LLLresponse}, their quadratic response is:
\begin{equation}\label{pc:tripletsusceptibility}
	\Pi_{\LLL}(\ii\omega,L) =-\rho_0g^2p_L(N)\left(\frac1{\Delta-\ii\omega} +\frac1{\Delta+\ii\omega}\right) =-\frac{2\rho_0g^2\Delta\,p_L(N)}{\Delta^2+\omega^2}.
\end{equation}
The term linear in frequency cancels. Expanding at small $\omega$ and $L(L+1)/N$ yields positive contributions:
\begin{equation}\label{pc:tripletstiffness}
	Z_{\tau,\LLL}=\frac{2\rho_0g^2}{\Delta^3},\qquad Z_{s,\LLL}=\frac{2\rho_0g^2R^2}{\Delta N}.
\end{equation}
There is also a direct way to see the absence of the LLL Berry term. In the color basis, $\mathsf M(\Phi)$ is real and symmetric, with negative off-diagonal entries $-h$. Its nondegenerate lowest eigenvector can be chosen real, normalized, and with positive components. Hence $u^T\partial_\tau u=0$, and the Berry connection vanishes throughout the region with one filled flavor band. The internal reflection $\mathsf R$ in \eqref{pc:permutations} also forbids $\Phi^*\partial_\tau\Phi-\Phi\partial_\tau\Phi^*$ in the interacting quadratic action. Unlike the polarized easy-plane doublet, this model therefore has no symmetry-allowed quadratic Berry obstruction at fixed flux.

Including the bare kernel, HLLs, and massive density fluctuations gives:
\begin{equation}\label{pc:pottseft}
	\cL_{\rm neutral}=Z_\tau|\partial_\tau\Phi|^2 +Z_s|\nabla\Phi|^2+r|\Phi|^2 +w(\Phi^3+\Phi^{*3})+u|\Phi|^4+\cdots.
\end{equation}
This is the conventional complex order parameter action with $S_3$ symmetry, the Landau--Ginzburg description of the three-state Potts model \cite{pc:ZiaWallace1975,pc:Amit1976,pc:ZinatiCodello2018}. The derivative expansion is valid below the charged and density gaps, with positive total $Z_\tau,Z_s$. Equations \eqref{pc:potential} and \eqref{pc:tripletstiffness} specify the fermion determinant contribution, while fully renormalized coefficients also contribute from the fluctuations of $\Phi$ itself (in particular, $r=0$ in a truncated potential is not an exact transition condition). The cubic interaction can produce a weak first order transition \cite{pc:JankeVillanova1997}. The precise nature of the fixed point does not follow from our calculation (the walking behavior was proposed in \cite{pc:GRZ2018}). The finite size evidence from \cite{pc:YangPotts} must constrain the dynamics of this EFT.

\subsection{The known two-body Hamiltonian}

Eliminating both Gaussian auxiliaries in \eqref{pc:tripletmother} gives:
\begin{equation}\label{pc:pottsout}
	\delta S=\frac12J_0K_\rho J_0-g^2J_Z^\dagger KJ_Z.
\end{equation}
Let $U$ denote the interaction used in \cite{pc:YangPotts}, written using the area-normalized densities of Section~\ref{sec:framework}. Choosing $K_\rho=2U$ and $g^2K=U$ and projecting to the LLL yields:
\begin{align}\label{pc:pottshamiltonian}
	H_3={}&\int\dd A_x\dd A_y\,U(x,y)
	\left[:n_0(x)n_0(y):-:n_Z(x)n_Z^\dagger(y):\right]
	-h\int\dd A\,[n_X+n_X^\dagger],\nonumber\\
	 n_T(x)={}&\psi^\dagger(x)T\psi(x),\qquad n_0=n_{\mathbf1}.
\end{align}
This is their equation (1), at $N_e=N$. In the interaction written here, the normal ordering contractions of the two terms cancel, since $ZZ^\dagger=\mathbf1$ and both terms use the same $U$.

For a finite sphere the equality is imposed on every retained harmonic:
\begin{equation}\label{pc:exactkernels}
	g^2K_L=U_L,\qquad (K_\rho)_L=2U_L, \qquad 0\le L\le N-1.
\end{equation}
The interaction in \cite{pc:YangPotts} has only the pseudopotentials $V_0,V_1$. Equivalently, $U$ is chosen so that its four-orbital coefficient tensor is:
\begin{equation}\label{pc:pairtensor}
	\mathcal V_{m_1m_2;m_3m_4} =\sum_{r=0}^{1}V_r\sum_M C^{(2s-r)M}_{s m_1,s m_2} C^{(2s-r)M*}_{s m_4,s m_3}.
\end{equation}
Using the Clebsch--Gordan to $3j$ relation gives exactly the tensor in their equation (4). This specifies the spatial kernel without a derivative truncation. The positive-covariance Gaussian formula above applies directly when $U_L>0$. For eigenvalues of either sign, diagonalize the real and imaginary color-density channels and use real or imaginary linear couplings with positive quadratic kernels, as in Section \ref{sub:twoorders}. Zero eigenvalues require no auxiliary. The inverse kernels $K^{-1}, K_\rho^{-1}$ in this exact representation can be spatially nonlocal. A local inverse kernel instead reproduces a low momentum expansion of $U$ and generally adds higher pseudopotentials.

\section{Topological order parameters: $SO(5)$, $O(4)$, and $Sp(n)$}\label{top:section}

The fuzzy sphere models in this section admit sigma model descriptions with a Wess--Zumino (WZ) term when their charged particles and fluctuations of the mass eigenvalues remain gapped. We start with Gaussian mothers that reproduce their interactions, then derive the sigma model in that regime. The constraint on the sigma model field therefore emerges in a low energy description and is not there in the UV: the HS field remains Gaussian in the UV. The calculation determines the topological term and the response to magnetic flux. We do not establish a real conformal fixed point, however: the infrared interpretations of the $SO(5)$ and $O(4)$ models remain open.

\subsection{The four-flavor Hamiltonian and its Gaussian mother}\label{top:clifford}

The constructions of Refs.~\cite{topIppoliti2018,topZhouSO52024,topChenPhase2024,topChen2024,topYang2026}
use four fermion flavors at half filling, $N_e=2N$, and interactions:
\begin{equation}\label{top:so5-hamiltonian}
	H_5=\int\dd A_x\dd A_y\left[U_0(x,y):n_0(x)n_0(y):-\sum_{I=1}^{5}U_I(x,y):n_I(x)n_I(y):\right].
\end{equation}
%Factors of $1/2$ in the contact couplings of some references are absorbed into the kernels $U_0,U_I$ in this convention.
Here $n_I=\psi^\dagger\Gamma^I\psi$ and $n_0=\psi^\dagger\psi$. The fields $\psi_a$, $a=1,\ldots,4$, are LLL projected, and the $4\times4$ matrices $\Gamma^I$ act on flavor space:
\begin{equation}\label{top:clifford-algebra}
	\{\Gamma^I,\Gamma^{I'}\}=2\delta^{II'},\qquad \tr(\Gamma^I\Gamma^{I'})=4\delta^{II'}.
\end{equation}
They are independent of the spacetime Dirac matrices. One choice is $\Gamma^{1,2}=\tau_{x,y}\otimes\mathbb I_2$ and $\Gamma^{3,4,5}=\tau_z\otimes(\sigma_x,\sigma_y,-\sigma_z)$, with $\Gamma^1\cdots\Gamma^5=\mathbb I_4$. Equal kernels $U_I=U$ give $SO(5)$ symmetry of the neutral bilinears. The charged fermions transform in the spinor representation of $Spin(5)=Sp(2)$; including their conserved electric charge, the faithful continuous group is $[U(1)\times Sp(2)]/\Z_2$.

Introduce five real scalars $\phi_I$ and a real density auxiliary $\rho$. With the convolution convention of Section~\ref{sub:twoorders}, the mother theory is:
\begin{equation}\label{top:so5-hs}
	S_5=\int\dd\tau\,\dd A\,\bar\Psi(\slashed D_A+\phi_I\Gamma^I+\ii\rho)\Psi+\frac14\sum_I\phi_IU_I^{-1}\phi_I+\frac14\rho U_0^{-1}\rho.
\end{equation}
The four components of $\Psi$ in this formula label flavors, each being a two-component Dirac field of charge one. The kernels are instantaneous. For positive kernel eigenvalues both Gaussian integrals are over real fields. Completing their squares gives $-\sum_I(\bar\Psi\Gamma^I\Psi)U_I(\bar\Psi\Gamma^I\Psi) +(\bar\Psi\Psi)U_0(\bar\Psi\Psi)$. The LLL projection, with the normal ordering corrections of Section~\ref{sub:twoorders}, is exactly \eqref{top:so5-hamiltonian}. Eigenvalues of the opposite sign are treated by interchanging real and imaginary linear couplings, as in the general HS construction. Derivative potentials are included directly through the harmonic eigenvalues of $U_I$; their inverse can be spatially nonlocal, as allowed in the reverse route.

At half filling the projected Hamiltonian also has the antiunitary particle--hole symmetry, written as $c_{ma}\mapsto\Omega_{ab}c^\dagger_{mb}$, $\ii\mapsto-\ii$ in a compatible symplectic flavor basis. This is the discrete symmetry used to identify spacetime parity in the spectral analysis of \cite{topZhouSO52024,topHuffman2026}. As in Section \ref{sub:isingmother}, this antiunitary particle--hole symmetry descends from the mother's flux-preserving $\widetilde{CT}$, with the flavor rotation $\Omega$ and the corresponding transformation of $\phi_I$ included. Its unitary counterpart involving spatial reflection descends from $\widetilde{CP}$. Both symmetries survive the $O(4)$ anisotropy considered below.

\subsection{The radial potential and the Wess--Zumino term}\label{top:radial}

Set $U_I=U$ and write $\bm\phi=r\bm n$, $r\geq0$, $\bm n^2=1$. The density saddle $i\rho_*\in\R$, if nonzero, is absorbed into the singlet source $m_0\bar\Psi\Psi$ that may be included. For a uniform $\bm\phi$, the flavor mass then has eigenvalues $+r,+r,-r,-r$. The two negative LLL states per orbital contribute $-2\rho_0r$ to the energy density. If $u=U_{L=0}>0$, the Gaussian HS action plus these LLL contributions give:
\begin{equation}\label{top:radial-LLL}
	V_{\LLL}(r)=\frac{r^2}{4u}-2\rho_0r,\qquad r_*=4u\rho_0,\qquad V_{\LLL}''(r_*)=\frac{1}{2u}>0.
\end{equation}
Thus the Gaussian field develops a preferred nonzero magnitude $|\phi|=r_*$ induced by the charged determinant. There is no need to constrain $\bm\phi$ to $S^4$: such a constraint emerges softly in the IR leading to an effective sigma model into $S^4$.

The HLL contributions change $r_*$ but preserve the qualitative picture. For $\lambda_j^2=j(Q+j)/R^2$ and $\rho_j=(Q+2j)/\mathcal A$, subtract the constant and quadratic ultraviolet terms into the vacuum energy and $u_R$. The remaining uniform potential is:
\begin{equation}\label{top:radial-HLL}
	V(r)=\frac{r^2}{4u_R}-2\rho_0r-4\sum_{j\geq1}\rho_j\left[\sqrt{\lambda_j^2+r^2}-\lambda_j-\frac{r^2}{2\lambda_j}\right].
\end{equation}
The factor four counts flavors. %For $r\ll\Delta_{\rm LL}$ the last term starts with $\frac12r^4\sum_j\rho_j/\lambda_j^3>0$. 
For $u_R>0$, we have:
\begin{equation}\label{top:radial-convexity}
	V''(r)=\frac{1}{2u_R}+4\sum_{j\geq1}\rho_j\left[\frac{1}{\lambda_j}-\frac{\lambda_j^2}{(\lambda_j^2+r^2)^{3/2}}\right]>0.
\end{equation}
With $V'(r)>\frac{r}{2u_R}-2\rho_0$ and $V'(0)=-2\rho_0$, this implies a unique solution to $V'(r_*)=0$, which is slightly smaller than the LLL value $4u\rho_0$. For $4u_R\rho_0\ll\Delta_{\rm LL}$, $r_*\approx 4u_R\rho_0$ is a reliable estimate.

These formulas evaluate the charged determinant, without the subsequent fluctuations of the HS fields. In the range where those corrections leave a nonzero radial scale $r_*$ and a charge gap, the remaining slow variable is $\bm n$, describing an effective sigma model at low energies. The uniform radial mode has no temporal kinetic term from the LLL alone (changing $r$ does not mix filled and empty flavors). HLL transitions generate a positive coefficient $Z_{r,\HLL}=\sum_{j\geq1}\rho_j\lambda_j^2/(\lambda_j^2+r^2)^{5/2}$ for $\frac12(\partial_\tau r)^2$. At leading order, we set $r=r_*$ (and estimate deviations away from $r_*$ later). %The positive $V''(r_*)$ above gives a finite static radial susceptibility $\frac1{V''(r_*)}$. A radial pole, if present, must also lie above the energies retained in the sigma model, so that the radial response is approximately instantaneous at low energy.

At fixed $r=r_*>0$ the charged part of the mother becomes:
\begin{equation}\label{top:so5-mother}
	\mathcal L_{5,\rm ch}=\bar\Psi(\slashed D_A+r_* n_I\Gamma^I)\Psi,\qquad \bm n^2=1.
\end{equation}
Its determinant gives a level-one WZ term by the familiar result of \cite[(34)--(39)]{topAbanovWiegmann2000}, or the Landau level derivation of \cite{topLeeSachdev2015,topHuffman2026}. With an orientation chosen so that the level is positive, the result is:
\begin{equation}\label{top:so5-effective}
	S_{\rm eff}[\bm n]=\int\dd\tau\,\dd A\left[\frac{\rho_\tau}{2}(\partial_\tau\bm n)^2+\frac{\rho_s}{2}(\nabla\bm n)^2+V_{\rm an}(\bm n)\right]+2\pi\ii\int_{B^4}\widetilde{\bm n}^{\,*}\omega_4+\cdots,
\end{equation}
where $\partial B^4$ is the 3D spacetime, a reference extension $\widetilde{\bm n}$ over $B^4$, $\widetilde{\bm n}\big|_{\partial B^4}=\bm{n}$ is understood, and
\begin{equation}\label{top:volume-form}
	\omega_4=\frac{1}{4!\,\Omega_4}\epsilon_{II'JKL}n_I\dd n_{I'}\wedge\dd n_J\wedge\dd n_K\wedge\dd n_L,\qquad \Omega_4=\frac{8\pi^2}{3},\qquad \int_{S^4}\omega_4=1.
\end{equation}
Here $V_{\rm an}$ is the angular potential caused by unequal interaction channels (it is constant in the $SO(5)$ invariant theory). For example, $U_{1,\ldots,4;L=0}=u$ and $U_{5,L=0}=u_5<u$ give, at fixed radius, $V_{\rm an}=\frac{r_*^2}{4}(u_5^{-1}-u^{-1})n_5^2$ when both are positive.

%The normalization of the topological term is checked as follows. The projector on the filled doublet is $P=(\mathbb I_4-\bm n\cdot\bm\Gamma)/2$, and the Clifford trace gives:
%\begin{equation}\label{top:projector-fourform}
%	-\frac{1}{8\pi^2}\tr\bigl[P(\dd P)^4\bigr]=\omega_4.
%\end{equation}
%This is a unit second Chern character. Its descent is the WZ functional above, whose level counts the number of four-component blocks $\Psi$ (we have only one).

The original HS field $\bm\phi$ takes values in $\R^5$, but the EFT derivative expansion is well-defined only when the charge gap, given by the mass scale $|\bm\phi|=r$ of fermions, remains large. Earlier we set the radial field to its approximate saddle value $r=r_*$. Thus at the very least, we want:
\begin{equation}
	E_{\rm EFT} \ll r_* \ll \Delta_{\rm LL}.
\end{equation}
Using LLL-generated $\rho_{\tau,\LLL}=\frac{\rho_0}{2r_*}$ and $\rho_{s,\LLL}=\frac{r_*}{2\pi}$ determined in the next subsection, the velocity is $v=\sqrt{\rho_s/\rho_\tau}$, and the EFT energies are $E_L=\frac{v}{R}\sqrt{L(L+1)}=2r_*\sqrt{L(L+1)/N}$. Thus the first inequality requires $L(L+1)\ll N$, with which we should also include $|\omega|\ll r_*$. What about fluctuations of $r$ away from $r_*$? A simple estimate can show that within the same EFT window, that is $|\omega|\ll r_*$ and $L\ll \sqrt{N}$, the fluctuations of $r$ remain parametrically small, while substantial excursions towards smaller $r$ have exponentially small probability. One may also worry that $r=r_*$ is not an exact saddle. Indeed, the EFT actually starts with the terms like $\cL=V(r) + \frac{\rho_0}{4r}(\partial_\tau {\bm n})^2 + \frac{r}{4\pi}(\boldsymbol{\nabla}\bm{n})^2+\dots$, so the radial equation of motion is not simply $V'(r_*)=0$; one may get deviations from $r_*$: $\frac{\delta r}{r_*} = \frac{(\partial_\tau \bm{n})^2}{8r_*^2} -\frac{(\boldsymbol{\nabla}\bm{n})^2}{8\pi\rho_0}+\dots$.
Under $|\omega|\ll r_*$, the first term is small, and the second term is small when the spatial variation length is large compared with the magnetic length, which again follows from $L\ll\sqrt{N}$.

Thus at low energies, the field $\bm\phi$ mostly remains within the spherical shell around $|\bm\phi|=r_*$. This justifies the standard effective low-energy description in terms of the sigma model into $S^4$. The excursions of field $\bm\phi$ outside the spherical shell manifest themselves as higher irrelevant terms in the $\bm{n}$ EFT. Furthermore, the shell having the topology of $S^4$ is what makes the WZ term possible. Indeed, if for whatever reason the values of $\bm\phi$ at low energies could explore the whole of $\R^5$, the WZ term would make no sense: it requires a nontrivial target topology. It is, however, consistent as a term in the low-energy sigma model into $S^4$.

\subsection{More on the response in magnetic flux}\label{top:flux}

For constant $\bm n$ the LLL energies are $\pm r_*$, each twice degenerate in flavor; the HLL energies are:
\begin{equation}\label{top:charged-spectrum}
	E_{j,\pm}=\pm\sqrt{r_*^2+\frac{j(j+Q)}{R^2}},\qquad j\geq1.
\end{equation}
The trace Berry curvature of the occupied doublet, whose projector is $P=(\mathbb I_4-\bm n\cdot\bm\Gamma)/2$, vanishes:
\begin{equation}\label{top:berry-zero}
	\ii\tr(P\dd P\wedge\dd P)=0.
\end{equation}
Indeed, traces of two Clifford matrices are symmetric in the $I$ indices, while traces of three distinct Clifford matrices vanish. This removes the term with one time derivative and no spatial derivatives. (The WZ term is related to $\tr(P(\dd P)^4)$, which is non-zero and proportional to $\omega_4$.)

For completeness, the quadratic response follows from transitions between the filled and empty doublets. Write
$\bm n=(\pi_1,\ldots,\pi_4,\sqrt{1-\bm\pi^2})$. The transition energy is $2r_*$. The quadratic part of $n_5$ supplies a local term that cancels the uniform static bubble, leaving, per component,
\begin{equation}\label{top:lll-kernel}
	\mathcal K_{\LLL}(\ii\omega,L)=2\rho_0 r_*\left[1-\frac{4r_*^2}{\omega^2+4r_*^2}p_L(N)\right].
\end{equation}
The quadratic action $\frac12\sum_{\omega LM,i}\pi^*_{i,\omega LM} \mathcal K\pi_{i,\omega LM}$, expanding at small $|\omega|$ and $L\ll\sqrt N$, gives:
\begin{equation}\label{top:lll-stiffness}
	\mathcal K_{\LLL}=\frac{\rho_0}{2r_*}\omega^2+\frac{2\rho_0r_*}{N}L(L+1)+\cdots,\qquad \rho_{\tau,\LLL}=\frac{\rho_0}{2r_*},\quad \rho_{s,\LLL}=\frac{r_*}{2\pi}.
\end{equation}
The temporal coefficient is simply the coefficient of $(\partial_\tau\bm n)^2/2$, and its positivity gives
the ordinary second order dynamics once the Berry term is absent. For the exact Gaussian kernel, also add the bare contribution $r_*^2(U_L^{-1}-U_{L=0}^{-1})/2$ to \eqref{top:lll-kernel}; HLLs would also contribute. Positive total coefficients give a relativistic two-derivative action after rescaling time, the usual $S^4$ NLSM with the level-one WZ term. This is the well-known model of deconfined quantum criticality \cite{TanakaHu2005,SenthilFisher2006,MaWang2020,Nahum2020,Nahum:2015vka,Wang:2017txt} that the fuzzy studies in \cite{topZhouSO52024,topChenPhase2024,topChen2024,topYang2026} were probing (see also \cite{Chester:2023njo} for a bootstrap study).

\subsection{The $O(4)$ deformation}\label{top:o4}

The model of \cite{topYang2026} weakens the fifth channel $u_5$ of \eqref{top:so5-hamiltonian}, keeping the other four equal to $u$. Its mother is therefore simply \eqref{top:so5-hs} with those unequal covariances. In the sigma model, when $0<u_5<u$, this produces the positive $n_5^2$ anisotropy described above. Below the fifth-component gap, $\bm n=(\bm N,0)$ with $\bm N\in S^3$. For a repulsive fifth channel $u_5<0$, one uses an imaginary HS coupling; the formula involving $u_5^{-1}$ above is only the attractive-channel description. Extending the sigma model to the $u_5<0$ part of the parameter range considered in \cite{topYang2026} (which includes their representative point $u_5/u=-0.75$) assumes that the charged and radial gaps persist.

Let us evaluate the $SO(5)$ WZ term after $\bm{n}(x)$ has been restricted to the equator $(\bm{N}(x),0)$. For that, we must extend $\bm{n}(x)$ into an auxiliary fourth (spacetime) coordinate, for which a convenient choice is $0\leq \alpha \leq \frac{\pi}{2}$ parameterizing an extension of $(\bm{N}(x),0)$ over the north pole: $\widetilde{\bm n}(x,\alpha)=(\sin\alpha\,\bm N(x),\cos\alpha)$. This gives:
%\begin{equation}\label{top:theta-latitude}
%	S_{\rm top}=\ii\theta(\alpha)\mathcal Q_3[\bm N],\qquad \theta(\alpha)=2\pi\left(\frac12-\frac34\cos\alpha+\frac14\cos^3\alpha\right),
%\end{equation}
\begin{equation}\label{top:o4-winding}
	S_{\rm WZ} = i\pi {\mathcal Q}_3[\bm N],\qquad \mathcal Q_3[\bm N]=\frac{1}{3!\,2\pi^2}\int\epsilon_{ijkl}N_i\dd N_j\wedge\dd N_k\wedge\dd N_l\in\Z.
\end{equation}
%The integral of $\sin^3\alpha$ over the northern hemisphere gives $\theta(\pi/2)=\pi$. 
Thus we get a winding number, or a theta-term. The EFT is the $O(4)$ sigma model at $\theta=\pi$, with the Berry cancellation and positive LLL kinetic coefficients inherited from the five-component theory. The numerical spectra in \cite{topYang2026} support approximate conformal behavior.

\subsection{The $Sp(n)$ Hamiltonian and its neutral mediators}\label{top:spn-hs}

We use $Sp(n)=USp(2n)$, with $n$ the symplectic rank and $N$ still the number of orbitals. The model of \cite{topZhouHe2025} has $2n$ fermion flavors, $a=1,\ldots,2n$, and filling $N_e=2mN$, $1\leq m<n$:
\begin{align}\label{top:published-spn}
	H_{Sp}={}&\int\dd A_x\dd A_y\left[U(x,y):n_0(x)n_0(y):\right.\nonumber\\
	&\left.\hspace{15mm}-\frac{V(x,y)}2\Omega_{ac}\Omega_{bd}\psi_a^\dagger(x)\psi_c^\dagger(y)\psi_d(y)\psi_b(x)\right],\qquad
	\Omega=\begin{pmatrix}0&\mathbb I_n\\-\mathbb I_n&0\end{pmatrix}.
\end{align}
Its continuous symmetry is $[U(1)\times Sp(n)]/\Z_2$. The tensor $\Omega_{ab}$ selects an antisymmetric flavor singlet $\Omega_{ba}\psi_a(y)\psi_b(x)$, so the orbital wavefunction is symmetric, and only the even pseudopotentials $V_0,V_2,\ldots$ contribute to the pairing interaction $V(x,y)$, while $U(x,y)$ has no such restriction.

To introduce neutral HS mediators, rewrite the interactions in terms of neutral bilinears $\psi^\dagger T\psi$, where $T$ are Hermitian $2n\times2n$ matrices acting on flavor indices. The identity matrix gives the total density $n_0$. Split the remaining traceless Hermitian matrices into the $\pm1$ eigenspaces of $T\mapsto\Omega^{-1} T^t\Omega$, and choose orthonormal bases $T_I^+$ and $T_I^-$ for these spaces:
\begin{equation}\label{top:bases}
	(T_I^+)^t\Omega=\Omega T_I^+,\qquad (T_I^-)^t\Omega=-\Omega T_I^-,\qquad \tr(T_I^\pm T_{I'}^\pm)=\delta_{II'}.
\end{equation}
Define $J_I^\pm=\psi^\dagger T_I^\pm\psi$ and the corresponding bilinears $\mathcal J_I^\pm=\bar\Psi T_I^\pm\Psi$ in the mother theory. The $T_I^+$ basis has $n(2n-1)-1$ elements and $J_I^+$ are the order parameters of \cite{topZhouHe2025}; the $T_I^-$ basis has $n(2n+1)$ elements and is the $Sp(n)$ adjoint.  The corresponding completeness identities are:
\begin{align}\label{top:completeness}
	\sum_I(T_I^+)_{ab}(T_I^+)_{cd}&=\frac12(\delta_{ad}\delta_{bc}+\Omega_{ac}\Omega_{bd})-\frac{\delta_{ab}\delta_{cd}}{2n},\nonumber\\
	\sum_I(T_I^-)_{ab}(T_I^-)_{cd}&=\frac12(\delta_{ad}\delta_{bc}-\Omega_{ac}\Omega_{bd}).
\end{align}
Subtracting them gives the exact normal-ordered decomposition:
\begin{equation}\label{top:spn-channel-hamiltonian}
	H_{Sp}=:n_0Wn_0:-\frac12\sum_I:J_I^+VJ_I^+:+\frac12\sum_I:J_I^-VJ_I^-:,\qquad W=U-\frac{V}{4n}.
\end{equation}

An explicit mother is:
\begin{align}\label{top:exact-spn-hs}
	S_{Sp}={}&\int\dd\tau\,\dd A\,\bar\Psi\left(\slashed D_A+m_0+X_IT_I^++\ii Y_IT_I^-+\ii\sigma\right)\Psi\nonumber\\
	&+\frac12\sum_IX_IV^{-1}X_I+\frac12\sum_IY_IV^{-1}Y_I+\frac14\sigma W^{-1}\sigma.
\end{align}
All mediators are neutral scalars. The singlet mass $m_0$ selects the required filling; its LLL contribution is constant in that filling sector. For positive $V$ and $W$, the displayed real Gaussian integrals give Eq.~\eqref{top:spn-channel-hamiltonian}. Other eigenvalue signs are handled channel by channel as in Eq.~\eqref{top:so5-hs}. We adopt the harmonic kernels of \cite{topZhouHe2025}.

For contact kernels the identity $\sum_{ab}:n_{ab}n_{ba}:=-:n_0^2:$, with $n_{ab}=\psi_a^\dagger\psi_b$, reduces the interaction to:
\begin{equation}\label{top:contact-fierz}
	\mathcal H_{Sp}^{\rm contact}=\left[U-\frac{n+1}{2n}V\right]:n_0^2:-V\sum_I:(J_I^+)^2:.
\end{equation}
For $n=2$, take $T_I^+=\Gamma^I/2$; this gives $(U-3V/4):n_0^2:-(V/4)\sum_{I=1}^5:n_I^2:$ and recovers the $SO(5)$ model of Section \ref{top:clifford}.

\subsection{The symplectic order parameter and its sigma model}\label{top:spn-mother}

The order parameter is $X=X_IT_I^+$, whereas $Y$ implements the adjoint channel in our HS decoupling of the Hamiltonian. The earlier $SO(5)$ Hamiltonian was already written in density and vector channels and did not require the adjoint channel (which can always be added or removed). For contact interactions, though, \eqref{top:contact-fierz} eliminates the adjoint channel at any $n$, leaving only $X$ and the density mediator.

We first compute the uniform potential from the Gaussian term and the LLL determinant, the analog of $V_\LLL(r)$ in the $SO(5)$ case. The eigenvalues of $X$ occur in pairs; at filling $N_e=2mN$, with $1\le m<n$, let $P(X)$ project onto the lowest $m$ pairs. Then the said potential is:
\begin{equation}\label{top:spn-potential}
	\mathcal V_{\LLL}(X)=\frac{\tr X^2}{4v}+\rho_0\tr[P(X)X],\qquad \rho_0=\frac{N}{\mathcal A}.
\end{equation}
Here $v=V_{L=0}/2$ for \eqref{top:exact-spn-hs}, or $v=V$ for \eqref{top:contact-fierz}. For $v>0$, minimizing over traceless $X$ gives a family of minima parameterized by projectors:
\begin{equation}\label{top:spn-radial}
	X_*(P)=-2\mu\left(P-\frac mn\mathbb I_{2n}\right),\qquad \mu=v\rho_0.
\end{equation}
The occupied and empty eigenvalues are respectively $-2\mu(1-m/n)$ and $2\mu m/n$. Their fluctuations cost $\tr(\delta X)^2/(4v)$ at fixed $P$, whereas their eigenspaces can rotate freely. Thus the analogue of fixing $r=r_*$ is fixing this spectrum of eigenvalues. We are free to rotate the projector $P$ that selects the eigenspaces, which is the analog of $\bm n$.

After eliminating the eigenvalue fluctuations, the same order parameter is parametrized by $X(x)=X_*(P(x))$, with
\begin{equation}\label{top:quaternionic-projector}
	P=P^\dagger=P^2,\qquad \tr P=2m,\qquad P^t\Omega=\Omega P.
\end{equation}
The projector is the NLSM variable here. Its orientations form:
\begin{equation}\label{top:grassmannian}
	\mathcal G_{m,n}=\frac{Sp(n)}{Sp(m)\times Sp(n-m)},\qquad
	\dim_{\R}\mathcal G_{m,n}=4m(n-m).
\end{equation}
For $n=2$, $m=1$, this is precisely the earlier parametrization: $X_*=\mu n_I\Gamma^I$ and $P=(\mathbb I_4-n_I\Gamma^I)/2$, with $\bm n^2=1$. Thus $P$ replaces $\bm n$ as coordinates for the order parameter.

Choosing the common mass shift $m_0+\ii\sigma_*=\mu(1-2m/n)$ gives:
\begin{equation}\label{top:spn-mother-action}
	\mathcal L_{\rm ch}=\bar\Psi\left[\slashed D_A+\mu(\mathbb I_{2n}-2P)\right]\Psi.
\end{equation}
The masses are now $\pm\mu$, and transitions cost $2\mu$. The adjoint saddle is $Y=0$, since $\tr(PT_I^-)=0$.

This same occupied projector determines the Berry response. Occupied states come in pairs $(u,\Omega u^*)$ whose Berry connections cancel, giving $\tr(P\dd P\wedge\dd P)=0$ at every filling. The term with one time derivative and no spatial derivatives therefore vanishes. Expanding the response~\eqref{top:lll-kernel} across the gap $2\mu$ gives:
\begin{equation}\label{top:spn-berry-inertia}
	\mathcal L_{\LLL}^{(2)}=
	\frac{\rho_0}{4\mu}\tr(\partial_\tau P)^2+
	\frac{\mu}{4\pi}\tr(\nabla P)^2,
\end{equation}
for $|\omega|\ll2\mu$ and $L(L+1)\ll N$ again.

The determinant also produces:
\begin{equation}\label{top:spn-wzw}
	S_{\rm WZ}[P]=2\pi\ii\int_{B^4}\widetilde P^{\,*}\nu_4,\qquad
	\nu_4=-\frac{1}{8\pi^2}\tr[P(\dd P)^4].
\end{equation}
Here $\widetilde P$ is the extension used to define the WZ term. Rotating just one occupied pair and one empty pair reduces the determinant to the four-flavor $SO(5)$ calculation; the other masses stay constant. This fixes the level to one, with $\int_{S^4}\nu_4=1$ on that embedded sphere~\cite{topZhouHe2025,topHuffman2026}. The WZ term includes spatial derivatives and survives the Berry cancellation.

The examples in Ref.~\cite{topZhouHe2025} are:
\begin{equation}\label{top:examples}
	\begin{array}{|c|c|c|c|}\hline
		(n,m)&\mathcal G_{m,n}&N_e&\dim_{\R}\mathcal G_{m,n}\\\noalign{\hrule height 1pt}
		(2,1)&S^4&2N&4\\\hline
		(3,1)&\mathbb H P^2&2N&8\\\hline
		(4,1)&\mathbb H P^3&2N&12\\\hline
		(4,2)&Sp(4)/(Sp(2)\times Sp(2))&4N&16\\\hline
	\end{array}
\end{equation}
Particle--hole conjugation sends $P\mapsto\mathbb I-P$ and $m\mapsto n-m$, reversing $\nu_4$. It preserves the filling only at $n=2m$, where this sign compensates spacetime orientation reversal (giving parity).

The reduction to $P$ assumes that charged, eigenvalue, and other mediator fluctuations remain gapped and that the total kinetic coefficients are positive. The resulting sigma model then has relativistic two-derivative dynamics, with numerical evidence for criticality following from \cite{topZhouHe2025}.

\subsection{The half-filled contact family}\label{top:qmc-family}

The Monte Carlo construction of Ref.~\cite{topHuffman2026} uses even $n$, $N_e=nN$, and contact interactions:
\begin{equation}\label{top:qmc-H}
	H_{\rm QMC}=\int\dd A\left[U:n_0^2(\bm r):-\frac{V_{\rm Q}}n\Delta^\dagger(\bm r)\Delta(\bm r)\right],\qquad \Delta=\frac12\Omega_{ab}\psi_a\psi_b.
\end{equation}
Since $4\Delta^\dagger\Delta=\Omega_{ac}\Omega_{bd}\psi_a^\dagger\psi_c^\dagger\psi_d\psi_b$, \eqref{top:contact-fierz} applies with $V=V_{\rm Q}/(2n)$:
\begin{equation}\label{top:qmc-fierz}
	\mathcal H_{\rm QMC}=W_{\rm Q}:n_0^2:-v_{\rm Q}\sum_I:(J_I^+)^2:,\qquad v_{\rm Q}=\frac{V_{\rm Q}}{2n},\quad W_{\rm Q}=U-\frac{n+1}{4n^2}V_{\rm Q}.
\end{equation}
The bilinears in this formula are exactly those defined around \eqref{top:bases}. A Gaussian mother is
\begin{equation}\label{top:qmc-HS}
	\mathcal L_{\rm QMC,HS}=\bar\Psi\slashed D_A\Psi+\sum_I\left[\frac{X_I^2}{4v_{\rm Q}}+X_I\bar\Psi T_I^+\Psi\right]+\frac{\sigma^2}{4W_{\rm Q}}+\ii\sigma\bar\Psi\Psi.
\end{equation}
For $W_{\rm Q}<0$ the density channel instead has cost $\sigma^2/(4|W_{\rm Q}|)$ and a real coupling $\sigma\bar\Psi\Psi$. The calculation above, with $m=n/2$, gives the sigma model on $Sp(n)/(Sp(n/2)\times Sp(n/2))$ at WZ level one. The study in \cite{topHuffman2026} finds evidence for a critical phase for $n\geq4$ among the even
ranks examined up to $n=16$, with correlators and spectra consistent with conformal symmetry. Their proposed $n$-flavor $SU(2)$ QCD$_3$ interpretation uses this numerical evidence and large-$n$ comparisons. On the contrary, our mother theory \eqref{top:qmc-HS} has no dynamical gauge field. We will construct gauge-theoretic mothers for \eqref{top:qmc-H} and other families in the upcoming work. In fact, a given fuzzy sphere Hamiltonian may admit multiple mother theories (some gauge-theoretic, some not), thus helping establish their IR dualities.

\section{Discussion and Outlook}\label{sec:discussion}

In this note, following the approach of \cite{MotherPRL}, we constructed and studied mother theories underlying the known fuzzy sphere Hamiltonians for the bosonic non-gauge CFTs. By ``non-gauge'' we mean that the CFT admits at least one RG flow described by a QFT that does not involve gauge field. Naturally, due to dualities, some of these models may also admit gauge-theoretic duality frames. We will study such gauge-theoretic descriptions, and how to construct their fuzzy sphere approximations, in the future work. We also chose to limit this paper to bosonic CFTs, simply because also including fermionic models would force us to squeeze too much information into one article, negatively affecting its readability.

In the examples above, each mother theory consists of charged massive Dirac fermions and neutral scalar mediators. The latter carry the CFT degrees of freedom, while fermions are restricted to the fixed filling sector needed to reproduce each Hamiltonian. One of the nice features of our approach is that this fixed filling is not imposed by hand: it emerges automatically as the appropriate Dirac vacuum on $S^2\times \R$ in the presence of magnetic flux, defined by filling all the negative-energy fermionic states. The fuzzy sphere Hamiltonian is obtained on the LLL by integrating out mediators. The neutral EFT for scalars, on the contrary, is computed in the same magnetic background by integrating out fermions, and it subsequently flows (after tuning relevant coupling) to the desired critical (or quasi-critical) point. In the easy-plane doublet, the occupied flavor state generates a Berry term and the saturation transition has $z=2$ Schrodinger symmetry. In the scalar, $O(n)$ rotor and three-state Potts models, a real occupied state or paired transitions remove this term and allow a relativistic EFT. In the symplectic family of models, the trace Berry term also vanishes, while a Wess--Zumino term is generated.

The exact HS identities and the derivative expansions answer different parts of this problem. A Gaussian integral fixes the projected LLL interaction at every retained harmonic. A charged determinant determines the leading neutral couplings, but its uniform saddle need not locate the transition of the interacting Hamiltonian due to further mediator fluctuations affecting physics. Where the published critical trajectory is known numerically on the fuzzy Hamiltonian side, we have identified the EFT and stated which steps have been established analytically.

There are several extensions and generalizations left for the future. Some of them are relatively straightforward and will be addressed in the upcoming publications, such as constructing mother theories for fermionic CFTs (this is mostly straightforward, except perhaps \cite{fer:Voinea} whose mother theory is not clear at the moment). Constructing gauge mother theories is less straightforward, but is both possible and highly desirable, so we plan to address soon. Some other future directions are mentioned below.

\subsection*{Extensions of the scalar constructions}
\paragraph{A four-flavor extension of the easy-plane mother.}
The easy-plane Berry obstruction suggests a new model. Couple the same $\Phi$ to two equally filled charged Dirac doublets with:
\begin{equation}\label{eq:balancedEP}
	\mathsf M_1=\mu\sigma^z+g(\operatorname{Re}\Phi\,\sigma^x+\operatorname{Im}\Phi\,\sigma^y),\qquad \mathsf M_2=\mu\sigma^z+g(\operatorname{Re}\Phi\,\sigma^x-\operatorname{Im}\Phi\,\sigma^y).
\end{equation}
Their occupied flavor states are complex conjugates, so their Berry terms cancel. Exchanging the doublets and conjugating $\Phi$ protects this cancellation at fixed flux. Gaussian mediator elimination gives:
\begin{equation}\label{eq:balancedEPReverse}
	\delta H=-g^2:J^+K_\Phi J^-: ,\qquad J^+=n_1^++n_2^-,\quad J^-=n_1^-+n_2^+.
\end{equation}
The even quadratic responses and positive determinant quartics add. These facts identify a possible $O(2)$ scalar EFT. It should be straightforward to verify it numerically.

\paragraph{A different $\Z_3$ interaction from the easy-plane doublet.}
The deformation suggested in \cite{MotherPRL} uses two charged flavors, rather than the three colors of Section~\ref{pc:potts}. Deforming linear Yukawa $g\Phi$ in the off-diagonal flavor mass by quadratic term,
\begin{equation}\label{pc:nonlineardoublet}
	A(\Phi)=g\Phi+\lambda\Phi^{*2},\qquad \mathsf M=m\sigma^z+A\sigma^-+A^*\sigma^+,
\end{equation}
is a relevant deformation that breaks the internal $U(1)$ to $Z_3$. The identity $|A|^2=g^2|\Phi|^2+g\lambda(\Phi^3+\Phi^{*3})+\lambda^2|\Phi|^4$ shows why a cubic scalar invariant appears in the EFT. Could this give another route to Potts? One issue is that the deformation starts at second order in $\Phi$ and does not change the quadratic Berry coefficient at the origin:
\begin{equation}\label{pc:epodd}
	c=-\operatorname{sgn}(m)\frac{\rho_0g^2}{8m^2}\ne0, \qquad S_B=c\int\dd\tau\,\dd A\, (\Phi^*\partial_\tau\Phi-\Phi\partial_\tau\Phi^*).
\end{equation}
Thus the cubic potential alone does not establish relativistic Potts (pseudo-)criticality. At first order in $\lambda$, eliminating the mediator gives:
\begin{equation}\label{pc:threeflip}
	\delta S=\lambda g^2\int\dd\tau\,\dd A\, \left[n^+(Kn^+)^2+n^-(Kn^-)^2\right]+O(\lambda^2).
\end{equation}
This is an interaction that flips three flavors and preserves magnon number only modulo three. It is not the two-body Hamiltonian. A contact covariance gives zero after LLL projection because $(n^+)^2=(n^-)^2=0$ at coincident points. A finite range covariance can avoid this cancellation: for the positive Green function $K=(r_\Phi-\kappa_\Phi\nabla^2)^{-1}$, $r_\Phi,\kappa_\Phi>0$, the matrix element from a filled down-flavor shell to the uniform three-flip state is proportional to $\int\dd A_x\dd A_y\dd A_z\,K(x,y)K(x,z)\det[\mathcal P(x_i,x_j)]$, which is strictly positive for $N\ge3$. Here $\mathcal P$ is the LLL projector; its Gram determinant is positive for generic distinct points. 

Oppositely polarized copies of $\Psi$ can cancel the Berry response, as in the proposal above. Determining the transition of the resulting model, including possible first order behavior, remains an open problem. Higher irrelevant clock deformations $A_q=g\Phi+\lambda_{q-1}(\Phi^*)^{q-1}$ can be examined in the same way, but their RG flow cannot be inferred from the invariant potential alone~\cite{pc:LouClock}. Whether the proposed direction is viable or not, it is also generally interesting to understand whether one can get any mileage out of including the higher-body interactions, such as \eqref{pc:threeflip}, in the fuzzy sphere Hamiltonians.

\paragraph{A compact mediator and the radial mode.}\label{pc:compact}
Another speculative proposal is to use a compact field $\theta\sim\theta+2\pi$ as a mediator. It removes the elementary radial field from $\Phi$, potentially giving a cleaner realization of the free scalar. For example, three charge-one Dirac flavors can have the mass matrix
\begin{equation}\label{pc:compactmass}
	\mathsf M_\theta=\begin{pmatrix} -m&g\cos\theta&g\sin\theta\\ g\cos\theta&m&0\\g\sin\theta&0&m \end{pmatrix},\qquad m>0.
\end{equation}
The action is $\bar\Psi(\slashed D_A+\mathsf M_\theta)\Psi$ plus an independently specified compact-scalar action. The flavor masses are $-\sqrt{m^2+g^2},\sqrt{m^2+g^2},m$. Filling the negative band therefore gives $N_e=N$ and a real occupied eigenvector with zero Berry connection. This provides a gapped charged sector coupled to a phase, but it is not yet a new purely fermionic fuzzy Hamiltonian. Eliminating $e^{\ii\theta}$ generates connected correlators of all orders. It remains to find a finite instantaneous Hamiltonian with the same neutral sector and a demonstrated reduction of unwanted scalar states.

There is also a curvature issue. After a velocity rescaling, a smooth compact-scalar action is $f^2(\partial\theta)^2/2$, with $[f^2]=1$. Its unpinned sphere frequencies are $\sqrt{L(L+1)}/R$, and its uniform
mode is a quantum rotor; they do not give the conformal scalar spectrum. A periodic pinning
\begin{equation}\label{pc:periodicpin}
	\delta\cL_R=\frac{f^2}{4R^2}(1-\cos\theta) =\frac{\varphi^2}{8R^2}-\frac{\varphi^4}{96f^2R^2}+\cdots, \qquad\varphi=f\theta
\end{equation}
gives the desired conformal mass locally. The anharmonic terms vanish in cylinder units when $f^2R\to\infty$, provided the theory is in its ordered phase and defects remain above the energies being studied. (The free $O(n)$ construction also suggests a projective mediator.)

\subsection*{Supersymmetric and fermionic mothers}\label{sub:fermionoutlook}
While we will study mother theories for fermionic CFTs in an upcoming work, there is one important question regarding supersymmetry. 
\paragraph{A supersymmetric mother for super-Ising.}
The tuned super-Ising critical point of the known fuzzy sphere model has numerical evidence for emergent supersymmetry~\cite{fer:ZGH}. A manifestly supersymmetric completion is not known. One possible flat-space starting point would use a real $\mathcal N=1$ superfield $\mathcal S=(\phi,\chi,F)$ and two charged complex scalar superfields $\mathcal Z_a=(B_a,\Psi_a,F_a)$, $a=0,1$, with canonical kinetic terms and real superpotential:
\begin{equation}\label{fer:SUSYcandidate}
	\mathscr W=\mathcal Z^\dagger(\mathsf M_0+h\mathcal S\sigma^x)\mathcal Z+\frac g6\mathcal S^3.
\end{equation}
The electric $U(1)$ is a flavor symmetry of the charged superfields. With $\mathsf M(\phi)=\mathsf M_0+h\phi\sigma^x$, eliminating the auxiliary component fields gives:
\begin{equation}\label{fer:SUSYpotential}
	V=B^\dagger\mathsf M(\phi)^2B+\frac12\left(\frac g2\phi^2+hB^\dagger\sigma^xB\right)^2.
\end{equation}
The same superpotential gives $\bar\Psi\mathsf M(\phi)\Psi$, the coupling of $\chi$ to $B^\dagger\sigma^x\Psi$ and its charge conjugate, and $g\phi\bar\chi\chi/2$. Thus both the even and odd currents have a supersymmetric origin. In a gapped supersymmetric charged vacuum, eliminating complete charged multiplets preserves supersymmetry and gives a neutral Wess--Zumino EFT, with its relevant superpotential terms retuned. The cubic interaction must be nonzero for the super-Ising target.

This action does not yet reproduce the published Hamiltonian on the fluxed sphere. It has two charged bosons, whereas that Hamiltonian has one, and its component potential contains additional interactions fixed by supersymmetry. Removing one boson or those interactions by hand would spoil the supersymmetric completion. Moreover, in the minimal $\mathcal N=1$ vector background, the gaugino variation is proportional to $F_{\mu\nu}\gamma^{\mu\nu}\varepsilon$~\cite{NishinoRajpoot2006}. A purely magnetic field gives invertible $\gamma^{12}$, so the usual bosonic monopole background alone preserves no such rigid supercharge. This is an obstruction for this background, not a no-go theorem for all supersymmetric mothers. Extended background multiplets and curved-space supersymmetry can supply further fields~\cite{ClossetDumitrescuFestucciaKomargodski2013}, but they also change the charged spectra and projected interactions. A supersymmetric flux background, the desired filling, and elimination of the extra charged multiplet must therefore be achieved together before \eqref{fer:SUSYcandidate} can be called a mother of the super-Ising model.

\subsection*{Relation to other fuzzy sphere calculations}\label{sub:scope}
Our catalogue concerns bulk Hamiltonians and their local continuum fields. The same regulator has been used to measure operator product coefficients and four-point functions~\cite{HuHeZhu2023,HanHuZhuHe2023}, construct conformal generators~\cite{FardelliEtAl2025,Fan2024Generators,FardelliEtAl2026OPE,EckWang2026}, and study entanglement and crosscaps~\cite{HuZhuHe2025F,DongEtAl2025Crosscap,RamirezCamascaEtAl2026}. The operator relation in Eq.~\eqref{eq:dictionary} supplies the corresponding EFT insertion, including contact terms. Locating the critical couplings and extracting these observables still require numerical input; available tools include FuzzifiED, quantum Monte Carlo, conformal perturbation theory, ground state energies, matrix product states, and reduced-basis calculations~\cite{Zhou2025FuzzifiED,HofmannEtAl2024,LauchliCPT,Wiese2026,HaoZhuQi2026,Baran2026}. Massive deformations can also be studied directly~\cite{FardelliEtAl2026IFT}. For a recent connection to matrix models, see \cite{Mizohata:2026rbp}.

Defects, impurities, and boundaries require additional local couplings or boundary conditions~\cite{HuEtAl2024Defects,ZhouEtAl2024GFunction,CuomoHeKomargodski2024,ZhouZou2025Surface,Dedushenko2024,SarmaEtAl2026,FengWang2026,YangHuChenZhu2026Defect,Tang:2026xny}. It must be possible to lift them to the mother theories from our catalogue. The noncommutative circle, $\mathbb{CP}^2$, and fuzzy three-sphere likewise have different orbital constructions~\cite{HanZhu2025Circle,MengZhaoZhuWu2026CP2,MengHuZhu2026S3}. These extensions, and the gauge sectors mentioned in the Introduction, require separate treatments.

The easy-plane example supplies a nonrelativistic continuum limit within the same method. Its dilute Bose gas is part of the broader study of nonrelativistic criticality~\cite{FisherWeichmanGrinsteinFisher1989,NishidaSon,Boisvert:2025hex,Boisvert:2026mgw}, including quantum Lifshitz theories~\cite{ArdonneFendleyFradkin2004,Antunes:2026jjf}, metallic order parameter dynamics~\cite{Hertz1976,Millis1993,OganesyanKivelsonFradkin2001}, and disordered quantum critical points~\cite{Fisher1995RandomTFIM}. These references describe different areas worth looking into.

\section*{Acknowledgments}
During the completion of \cite{MotherPRL}, the author benefited from conversations and/or correspondence with Yin-Chen He and Zheng Zhou. This current article is a significantly expanded version of the talk presented at Strings-2026 on July 7, 2026.

\paragraph{Statement on the AI use.} The results were obtained by the author, with the assistance of ChatGPT for calculations and rapid exploration of ideas, over a period from January to October 2026. All the results are verified by the author, who bears full responsibility for everything in this work. The manuscript based on the research materials collected over this period of time was prepared with the assistance of ChatGPT Astra, and subsequently heavily edited and almost entirely rewritten by the author.

\appendix
\section{Charged response and LLL form factors}\label{app:spectral}

The following formulas underlie the Ising, $O(n)$, Potts, and $Sp(n)$ responses used above. Let $h_*$ be the charged one-body Hamiltonian at a uniform background, with eigenstates $|p\rangle$, energies $\epsilon_p$,
and occupations $n_p=0,1$. Indices $I,I'$ label real mediator components; $p,q$ include orbital and flavor labels. For a local dependence $h(X)$, the vertex of an area-normalized harmonic is:
\begin{equation}\label{eq:vertexspectral}
	V^{I,LM}_{pq}=\int\dd A\,Y_{LM}(x)u_p^\dagger(x)(\partial_Ih)_*u_q(x).
\end{equation}
Use the Fourier convention of Section~\ref{sub:flux} and define the quadratic kernel by:
\begin{equation}\label{eq:quadraticConvention}
	\Gamma_f^{(2)}=\frac12\int\frac{\dd\omega}{2\pi}\sum_{LM}X^I_{LM}(\omega)^*\Pi_{II'}(\ii\omega,L)X^{I'}_{LM}(\omega).
\end{equation}
Expansion of $-\Tr\log(\partial_\tau+h)$ gives the bubble (see also Supplemental Material of \cite{MotherPRL}):
\begin{equation}\label{eq:spectralsum}
	\Pi^{\rm bub}_{II'}(\ii\omega,L)=\frac1{2L+1}\sum_{M,p,q}\frac{n_p-n_q}{\ii\omega+\epsilon_p-\epsilon_q}\,V^{I,LM}_{pq}(V^{I',LM}_{pq})^*.
\end{equation}
Rotational invariance permits averaging over $M$. The order of the vertices in this equation follows the stated quadratic action convention. For nonlinear $h(X)$, the second variation also contributes:
\begin{equation}\label{eq:seagullspectral}
	\Pi^{\rm sg}_{II'}(L)=\frac1{2L+1}\sum_{M,p}n_p\int\dd A\,|Y_{LM}|^2u_p^\dagger(\partial_I\partial_{I'}h)_*u_p.
\end{equation}
The full charged contribution is $\Pi^{\rm bub}+\Pi^{\rm sg}$, with the same regulator and vacuum subtraction as the determinant. 

For a diagonal entry, occupied--empty transitions combine to:
\begin{equation}\label{eq:evenspectral}
	\Pi^{\rm bub}_{II}(\ii\omega,L)=-\frac2{2L+1}\sum_{M,p\in{\rm occ},q\in{\rm emp}}\frac{\Delta_{qp}|V^{I,LM}_{pq}|^2}{\omega^2+\Delta_{qp}^2},\qquad \Delta_{qp}=\epsilon_q-\epsilon_p>0.
\end{equation}
Its $\omega^2$ coefficient is nonnegative, being a sum of $2|V|^2/\Delta^3$. Off-diagonal entries additionally contain terms odd in $\omega$, proportional to imaginary products of distinct vertices. They produce the imaginary Euclidean Berry action and must be kept. The kernel obeys $\Pi_{II'}(\ii\omega,L)=\Pi_{I'I}(-\ii\omega,L)$. Equation~\eqref{eq:Berryprojector} gives an independent
way to compute its leading Berry term.

Let $P_0$ now denote the projector onto the single-particle LLL. The orbital trace common to these responses is:
\begin{equation}\label{eq:formfactappendix}
	\Tr_{V_s}\!\left[(P_0Y_{LM}P_0)^\dagger(P_0Y_{LM}P_0)\right]=\rho_0p_L(N)=\frac N{\mathcal A}\frac{(N-1)!N!}{(N-L-1)!(N+L)!},\qquad 0\le L<N.
\end{equation}
The product formula in Eq.~\eqref{eq:LLLresponse} gives:
\begin{equation}\label{eq:pLlog}
	\log p_L(N)=-\frac{L(L+1)}N+O\!\left(\frac{L^2(L+1)^2}{N^3}\right),\qquad L\ll N.
\end{equation}
A low-order Taylor expansion of $p_L$ itself requires the stronger condition $L(L+1)\ll N$. Multiplying this orbital trace by the actual flavor transition matrix elements and using their excitation energies then yields the responses in the main text.

%\clearpage
\phantomsection
\addcontentsline{toc}{section}{References}
\begingroup
\small
\setlength{\bibsep}{2pt}
\interlinepenalty=10000
\bibliographystyle{utphys}%unsrtnat}
\bibliography{non_gauge_mother_theories}
\endgroup
\end{document}